\documentclass[a4paper]{jps-cp}

\usepackage{footmisc}
\usepackage{bm}

\usepackage{txfonts}
\usepackage{url}
\usepackage{multicol}
\usepackage{capt-of}

\usepackage{booktabs}
\usepackage{longtable}
\usepackage{ltcaption}

\usepackage{subfigure}  

\newcommand\newblock{\hskip .11em plus .33em minus .07em}
\newcommand\figref[1]{Fig.~\ref{#1}}
\newcommand\tabref[1]{Table~\ref{#1}}
\newcommand\secref[1]{Section~\ref{#1}}

\title{%
  Bitcoin's Crypto Flow Network}

\author{%
  Yoshi \textsc{Fujiwara}$^{1}$ %
  and Rubaiyat \textsc{Islam}$^{1}$}

\inst{%
  $^{1}$Graduate School of Information Science, University of Hyogo, Kobe 650-0047, Japan}

\email{yoshi.fujiwara@gmail.com}

\recdate{May 31, 2021}

\abst{%
  How crypto flows among Bitcoin users is an important question
  for understanding the structure and dynamics of the cryptoasset
  at a global scale. We compiled all the blockchain data of Bitcoin
  from its genesis to the year 2020, identified users from anonymous
  addresses of wallets, and constructed monthly snapshots of networks
  by focusing on regular users as big players. We apply the methods of
  bow-tie structure and Hodge decomposition in order to locate the
  users in the upstream, downstream, and core of the entire crypto
  flow.  Additionally, we reveal principal components hidden in the
  flow by using non-negative matrix factorization, which we
  interprete as a probabilistic model. We show that the model is
  equivalent to a probabilistic latent semantic analysis in natural
  language processing, enabling us to estimate the number of
  such hidden components. Moreover, we find that the bow-tie structure
  and the principal components are quite stable among those
  big players. This study can be a solid basis on which one can
  further investigate the temporal change of crypto flow,
  entry and exit of big players, and so forth.\\[5pt]
  \textsl{Presented at the conference ``Blockchain in Kyoto 2021'',
  and forthcoming in the JPS Conference Proceedings.}
}

\kword{Bitcoin, cryptoasset, bow-tie structure, Hodge decomposition,
  non-negative matrix factorization, latent Dirichet allocation, complex network}

\begin{document}
\maketitle

\section{Introduction}\label{sec:intro}

Cryptoasset or cryptocurrency is essentially a digital ledger to
record transactions between creditors and debtors, just like
money. The digital system is based on a collection of non-centralized
ledgers, called blockchain, which contains all the historical record
of transactions among anonymous users. Today there are many
cryptoassets being exchanged in markets with fiat currencies and also
with each other. The market capitalization is so huge in total ranging from
one to a few trillion USD, and highly volatile potentially having a
big impact even on asset markets and prices of non-crypto at a global
scale.

In this paper, we study Bitcoin, the largest one dominating nearly
half of the market capitalization at the time of writing. We attempt
to understand the flow of crypto as a complex network comprising of the
users as nodes and the crypto flow as links. There are a number of
studies from such a viewpoint of complex network on cryptoassets.
See %
\cite{reid2013aab,ober2013sab,ron2013qaf,kondor2014drg,kondor2014iin,alvarez2014nca,baumann2014ebn,fleder2015btg,maesa2016ubb,lischke2016abn,akcora2017bgp,bartolettei2017gfb,cachin2017tgm,cazabet2017tbu,maesa2017dda,ranshouts2017epm} %
for example, and references therein.

Specifically, in this paper, we focus on ``big players'' who are
defined as persistently appearing users, likely to be involved in
transactions of high frequencies and large amounts, and address the
following questions. First, it is important to identify the users in
the upstream, downstream, or core in the entire crypto flow. We shall
examine the so-called ``bow-tie'' structure of the network of those
big players to classify the location of the crypto flow based on the
binary relationship of links. Second, we measure the location in a
more quantitative way by using the information of flow along the links
in the combinatorial method of Hodge decomposition. Third question is
to extract ``principal components'' hidden in the entire crypto flow
so as to uncover a certain number of latent factors or components.
Fourth, because the network is changing in time, what can one say
about the stability of the crypto flow?

In \secref{sec:data}, we describe our dataset of Bitcoin to identify
the anonymous addresses representing wallets into users. Then we
define regular users as big players and construct networks.
In \secref{sec:nw_prop}, we perform the bow-tie analysis to locate
users in the stream of crypto flow. In \secref{sec:hodge}, we use the
method of Hodge decomposition to quantify the location of users.
In \secref{sec:nmf}, we introduce non-negative matrix factorization as a method of matrix
decomposition to reveal principal components hidden in the flow. We
shall show in \secref{sec:nmf_prob} that the method can be interpreted as a probabilistic
model, from which one can estimate the number of components. In \secref{sec:nmf_res},
we find about a dozen of principal components among several hundred big
players, and show that the temporal change of the network is quite
stable. In \secref{sec:discuss}, we discuss about several
aspects, being worth further investigation, and conclude in \secref{sec:summary}.
We add Appendix~\ref{sec:identiy} for the identities (actual names,
business, and so forth) of selected users.
Appendix~\ref{sec:adj} illustrates the networks in adjacency matrices.
Appendix~\ref{sec:lda} briefly summarizes the above mentioned
probabilistic model in relation to latent Dirichlet allocation,
from which the estimation on the number of components is done
in Appendix~\ref{sec:optnum}.

\section{Data}\label{sec:data}

We employ the dataset of all transactions recorded in the Bitcoin
blockchain from the genesis block (first block issued on January 9,
2009) until the block of height 63,299 (inclusive; issued on June 4,
2020). Each transaction is a transfer of a certain amount of BTC
(monetary unit of Bitcoin) from one or more addresses to others as we
will see shortly. We call such a transfer of BTC as crypto flow. An
address is something like a wallet possessed by a user who can be an
individual or, more frequently today, an agent in the business of
exchanges, services, gambling, and so forth.

In the dataset, total number of transactions was 1.38 billion, while
the number of different addresses was about 657 million. To study
crypto flow of Bitcoin, one needs to know users, rather than the
addresses. However, it is not straightforward to identify users from
addresses because of the very nature of anonymity inherent in the core
technology of blockchain. See \cite{oreilly2017mb} for
technical details.

Let us employ a simple but useful method to identify users from
addresses to construct a giant graph comprising of nodes as users and
edges as crypto flow. We shall see that more than 60\% of the
addresses can be identified with users. Additionally we will see that
a number of users can be revealed with their actual names, types of
business, and sometimes geographical location at a global scale. Then we
define regular users as big players in order to focus on a subgraph
comprising of frequently appearing users who are involved in crypto
flow with huge amounts of BTC. The subgraph will be studied in the
subsequent sections.

\subsection{Identification of Users from Addresses}\label{sec:data_users}

Consider an example of such a transaction (TX) that one day Alice
transferred 1 BTC to Bob:
\begin{equation}
  \label{eq:tx1}
  \text{TX}_1: \{a_1,a_2\}\rightarrow\{a_{123},a_1\}\ ,
\end{equation}
where the addresses $a_1$ and $a_2$ belong to Alice, while $a_{123}$
belongs to Bob. Alice needs more than one address as input of
$\text{TX}_1$, because a single one was not sufficient to
fulfill the amount of 1 BTC. Output of $\text{TX}_1$ includes $a_1$
representing the change. Another day, Alice did another transaction:
\begin{equation}
  \label{eq:tx2}
  \text{TX}_2: \{a_1,a_3\}\rightarrow\{a_{45},a_3\}\ ,
\end{equation}
where the address $a_3$ also belongs to Alice. Obviously, multiple
addresses, if and only if they appear in an input of a transaction,
belong to the same user, namely her wallets. As a consequence from
both of \eqref{eq:tx1} and \eqref{eq:tx2}, it follows that $a_1$,
$a_2$, $a_3$ can be identified to belong to the same user. Note that
$a_2$ and $a_3$ did not appear in any record of transactions,
$\text{TX}_1$ and $\text{TX}_2$. By looking at all the transaction in
the history, one can identify many addresses with users.

This simple but useful method to identify users from addresses was
proposed by \cite{reid2013aab} and has been extensively used in the
literature (see \cite{kondor2014drg,kondor2014iin} and the data of
\cite{hungary2020data,walletexplorer}, for example). We implemented
an efficient algorithm of this method to process the above mentioned 657
million addresses in the 1.375 billion transactions. We found that 402
million addresses (more than 60\%) can be identified to obtain 60
million different users (denoted as \textit{type A}). The rest of 255
million addresses, which never appeared as multiple inputs in any
transactions, are regarded as different users (\textit{type B}). This
procedure results in 315 million users all together.

\begin{figure}[tb]
  \centering
  \includegraphics[width=0.60\linewidth]{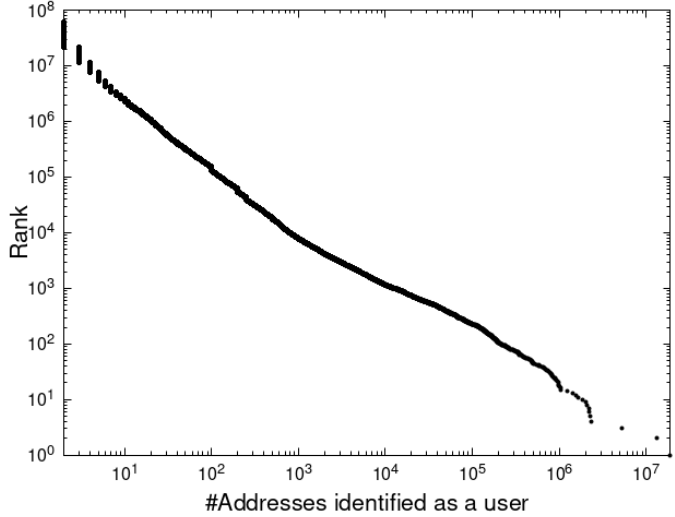}
  \caption{%
    Rank-size plot for the number of addresses identified with users.
    Size (horizontal) is the number of addresses identified with
    users of type A (a group of addresses appearing in multiple inputs
    of transactions, so identified as user; see main text).  Rank
    (vertical) is the descending order of the size. Maximum size is
    nearly 20 million, while minimum is 2 and average is 6.7.}
  \label{fig:uid-num_addr}
\end{figure}

\figref{fig:uid-num_addr} shows a rank-size plot, in which the size is
the number of addresses identified with users of type A, and the rank
is its descending order. The minimum size is 2 by construction, while
the average is 6.7. It is interesting to observe that the rank-size
plot is highly skewed with the maximum size being nearly 20 million!
We labeled all the users of type A in a sequential user ID
corresponding to the rank (e.g. \texttt{0000012345}), while users of
type B is labeled with its address
(e.g. \texttt{3CjqmbuRA1LEWmLHiWoSWHcWuTEVPfU24P}).

\begin{figure}[tb]
  \centering
  \includegraphics[width=0.99\linewidth]{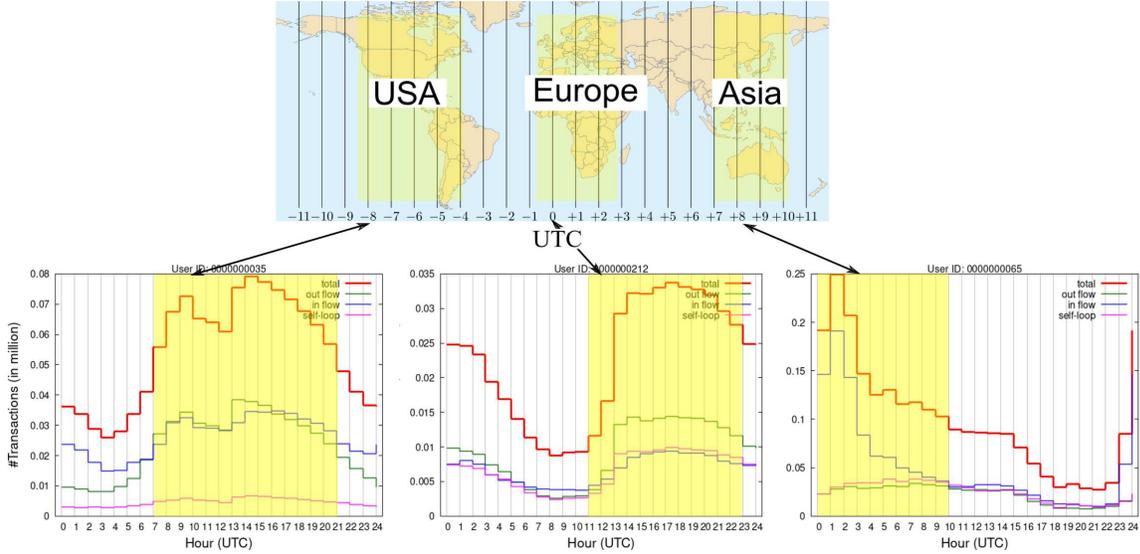}
  \caption{%
    Schematic diagram for the intra-day activities of users.
    Top is a world map with UTC (Coordinated Universal Time).
    The bottom three plots depict the number of transactions
    involving three users. Each plot shows the numbers of
    transactions of out-flow (green), in-flow (blue), self-loop (magenta),
    and their total (bold red), where each number is accumulated during
    the year 2019. Out-flow, in-flow, and self-loop are respectively
    the transaction in which the user is its source, destination, and both
    of them simultaneously. A peak of the red line shows the most
    active time, corresponding to the daytime of USA, Europe, and Asia
    (left to right). Arrows indicate noon of specific locations.}
  \label{fig:hist_tx_utc_freq}
\end{figure}

A bunch of transactions is packed into a block in the blockchain. Each
block has a timestamp of its birth, which is UTC (Coordinated
Universal Time) when the block was mined or born as an empty ledger to
be filled with transactions. We associate the time with the
transactions contained in the block. Each block is mined in a mean
time of 20 minutes or longer, as is well known, so it should be
understood that by time we mean a rough estimate on when the
transaction was made within an accuracy of an hour or so.
One can use this information of the timestamp to obtain the intra-day
activities of users. \figref{fig:hist_tx_utc_freq} is a schematic
diagram for typical three users, presumably located in Asia,
Europe including Africa, and USA.

Moreover, with the help of laborious investigation by
\cite{walletexplorer}, it is possible to unravel the identity of users
of type A with actual names, types of business, and sometimes
geographical location for many users. Appendix~\ref{sec:identiy} is
the summary of the unraveled identity. \tabref{tab:walletexplorer_cat}
gives the classification of business into exchanges, gambling, pools,
services and others. Nearly one third in the list are exchanges.
\tabref{tab:walletexplorer_country} is the list of countries of those
exchanges, where China, UK, and USA are dominant, followed by Canada,
Australia, Brazil, Singapore, and Russia. In fact, as found in the top
of \tabref{tab:walletexplorer}, the user ID \texttt{0000000000}
corresponding to the maximum size in \figref{fig:uid-num_addr} is
actually \texttt{Bit-x.com} and \texttt{Xapo.com}, the former of which
is an agent of exchange in South Africa.

Exchanges are a typical category of ``big players'' in the sense that
they actually hold a huge number of individuals and agents as
customers resulting in a large number of transfers. As a matter of
fact, the daily number of transfers has an interesting weekly pattern.
There is significantly less activity in weekends than in weekdays as
found in recent data of Bitcoin\footnote{%
  Cryptoasset of XRP has a similar weekly pattern according to
  \cite{akky2021pc}. In the early history of Bitcoin, the weekly
  pattern was quite the opposite; more active in weekends,
  presumably because individuals were big players at that time.} %
(see our previous study \cite{islam2019aoa,islam2020uif}).
Such a weekly pattern implies that those institutional agents are
dominant in the entire flow of crypto.


\subsection{Crypto Flow Network}\label{sec:data_nw}

Now all the transactions among addresses are converted into transfers
from users to users, each of which has the following information:
\begin{itemize}\setlength{\itemsep}{0pt}
\item user of source $s$,
\item user of destination $d$,
\item amount of Bitcoin transferred from $s$ to $d$, i.e. $s\rightarrow d$,
\item UTC time of transfer (from the block containing the transaction).
\end{itemize}

\begin{figure}[tb]
  \centering
  \includegraphics[width=0.75\linewidth]{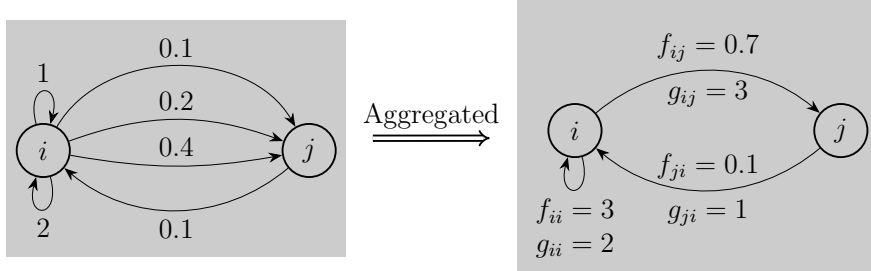}
  \caption{%
    Aggregation of transactions to construct crypto flow network.
    \textit{Left}: During a given period of time, for a pair of users
    $i$ and $j$, three transfers for $i\rightarrow j$ with 0.1, 0.2,
    0.4 in BTC, one of $j\rightarrow i$ with 0.1, and two for
    $i\rightarrow i$ with 1 and 2. \textit{Right}: After aggregation,
    one has $i\rightarrow j$ with frequency $f_{ij}=3$ and amount of
    flow $g_{ij}=0.7$. Similarly, $j\rightarrow i$ with $f_{ji}=1$
    and amount of flow $g_{ji}=0.1$, and $i\rightarrow i$ with
    $f_{ii}=2$ and $g_{ii}=3$.}
  \label{fig:aggregate}
\end{figure}

During a certain period of time $T$, for a pair of users, there can be
more than one transfer as depicted in \figref{fig:aggregate} (see the
left-hand side). In this example, there are three transfer of crypto
for $i\rightarrow j$, one for $j\rightarrow i$, and also two for
$i\rightarrow i$. The last case of self-loop is possible, because one
can receive a change in a transaction, and also because different
addresses are possibly identified with a user, like an exchange.
Given a time-scale $T$, it would be reasonable to aggregate these
transactions as shown in the right-hand side of
\figref{fig:aggregate}.

After the aggregation, one has a network comprising of nodes as
users and edges with direction given by the transfer of crypto,
frequency and amount of transfer occurred during the period of time.
Let us denote the following variables which represent the strength
or weight of each edge by
\begin{description}\setlength{\itemindent}{5em}
\item[frequency] $f_{ij}\equiv$ frequency of transfers for
  $i\rightarrow j$,
\item[amount of flow] $g_{ij}\equiv$ amount of total transfers for
  $i\rightarrow j$.
\end{description}

Regarding the time-scale $T$ for aggregating the transactions and the
epoch to select in the historical data, we choose one month and the
calendar year of 2019. By examining the time-series for the daily
number and amount of transactions, we assumed that the period of one
month is adequate to study the stability and temporal change of the
crypto flow. Shorter period may lead to a trivial result for the
stability and could be insufficient to detect the temporal change, if
any change is present. Also longer period would be misleading due to
non-equilibrium nature of the system. The year 2019 was chosen as the
epoch, in which one does not see violent bubble or crush in the price.

The number of all the users is huge, 315 million users in total.
Fortunately, however, it is not necessary to include all of them,
because most of them do not appear frequently. In the next section, we
shall extract only a tiny part of the network by focusing on the
``regular users'' who appeared everyday during the specified period.

\subsection{Regular Users as Big Players}\label{sec:data_reg}

\begin{figure}[tb]
  \centering
  \includegraphics[width=0.78\linewidth]{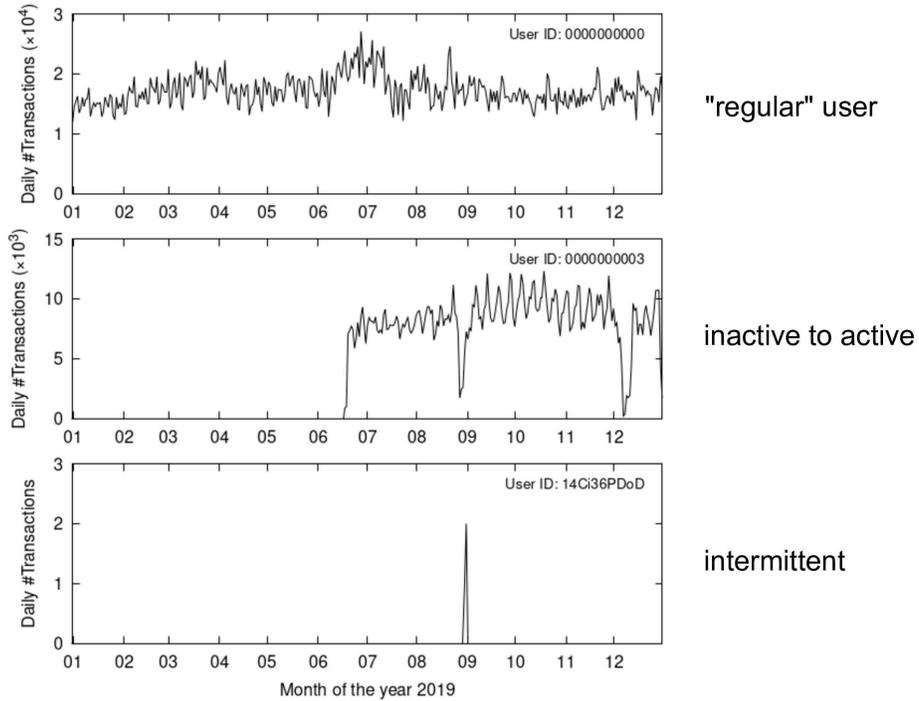}
  \caption{%
    Illustrative examples for the activities of users. Each plot
    depicts the daily number of transactions, in each of which the
    user is either source or destination of the transaction
    for the period of year 2019.
    Self-loops (case of the same source and destination) are excluded.
    Top is a ``regular'' user appearing every day. The middle user
    became active from being inactive, while the bottom one has
    intermittency in its activity. We focus on regular users
    in this paper.}
  \label{fig:def_regular}
\end{figure}

\begin{figure}[h]
  \centering
  \includegraphics[width=0.48\linewidth]{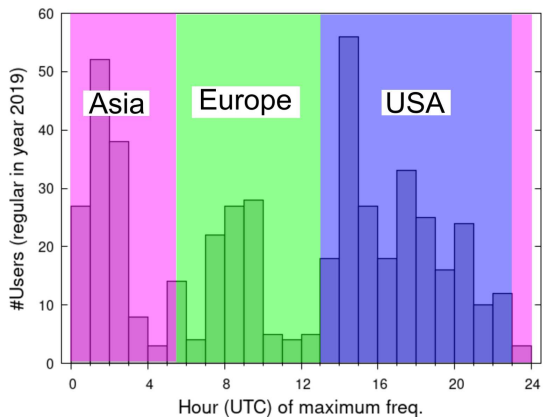}
  \caption{%
    Activities of regular users, defined by each user's peak of
    transactions, in UTC. Shaded region corresponds to the daytime
    of Asia, Europe, and USA (left to right).}
  \label{fig:stat_hist_tx_utc_freq}
\end{figure}

For our purpose in this paper, it is sufficient to focus on the crypto
flow with high frequency and big amount of Bitcoin, because infrequent
and/or small amount of flows is obviously unimportant for the
understanding the entire flow. In other words, it suffices to focus on
``big players'' who are playing some dominant role in the game of
crypto flow. It would be possible to define such a big player in
different ways. In this study, we define it by looking at how
persistently the user appears in transactions during our specified
period of time.

\figref{fig:def_regular} depicts examples of users who appear in
different numbers of transactions on daily basis. The user of the top
case is persistent in committing transactions with other users, which
can be labeled as a ``regular user''.  The middle case changes the
persistency from being inactive with no transaction with anyone else
to being active in an abrupt way. The bottom case has little activity,
just a few transactions on particular days having a strong
intermittency.

We define \textit{regular users} as those appearing everyday during
the period of one year in 2019, and use them as big players. The
number of regular users was 479. Then we construct a subgraph in each
month, which is comprised of the regular users as nodes and the crypto
flow as links, the latter of which are aggregated as described in
\figref{fig:aggregate}. Thus we have 12 snapshots of such subgraphs,
each corresponding to each month in the year, from January to
December. Summarizing the processing of the whole dataset, we
constructed the snapshots of networks denoted by $G_t=(V_t,E_t)$,
where $t$ is the month, $V_t$ is the set of regular users, $E_t$ is
the set of links among them, each having the frequency and amount as
depicted in \figref{fig:aggregate}.

We add \figref{fig:stat_hist_tx_utc_freq} showing the histogram for
the UTC time of highest activities of all the regular users in the
year of 2019. This information gives geographical locations of those
regular users.

\section{Analysis of Crypto Flow Network}\label{sec:nw}

\subsection{Basic Properties of Network}\label{sec:nw_prop}

For each month $t$, we constructed a network denoted by
$G_t=(V_t,E_t)$ as described in the preceding section.
\tabref{tab:nw_basic} is the summary of basic properties
of the networks in the year 2019. $V_t$ corresponds to the regular users
in each month, the number of which is shown by the column $|V_t|$.
The column $|E_t|$ is the number of edges, namely the number of
different crypto flow from one user to another or to itself (self-loops
as shown in parentheses). Most of the users have self-loops.
Temporal change of the network causes the changes of $V_t$ and $E_t$.
The column of $|V_t\cap V_{t+1}|$ is the number of users that
are common to successive months in their appearance. One can see that
most of the users are appearing successively. The same is true
for the edges as shown in the column of $|E_t\cap E_{t+1}|$.
In other words, the network is not changing drastically in terms of
the entry and exit of nodes and edges during the time-scale of months.

\begin{table}[tbh]
  \centering
  \caption{%
    Summary of basic properties of the networks in the year 2019, January to December.}
  \begin{tabular}{ccccccc}
    \toprule
    $t$ & $|V_t|$ & $|E_t|$ & $|V_t\cap V_{t+1}|$ & $|E_t\cap E_{t+1}|$ & GWCC & GSCC/IN/OUT/TE \\
    \midrule
    01 & 470 & 17,215(408) & 468 & 13,313 & 468(3) & 327/24/113/4 \\
    02 & 470 & 16,658(407) & 468 & 13,357 & 468(3) & 322/20/120/6 \\
    03 & 473 & 17,618(408) & 471 & 13,942 & 470(4) & 327/25/113/5 \\
    04 & 473 & 17,691(415) & 470 & 13,960 & 469(5) & 336/16/115/2 \\
    05 & 472 & 17,903(416) & 471 & 14,012 & 468(5) & 330/23/112/3 \\
    06 & 471 & 17,787(412) & 469 & 13,705 & 467(5) & 318/23/124/2 \\
    07 & 472 & 17,292(410) & 468 & 13,108 & 470(3) & 333/21/110/6 \\
    08 & 468 & 16,534(402) & 467 & 12,628 & 466(3) & 321/18/123/4 \\
    09 & 470 & 16,288(407) & 468 & 12,652 & 466(5) & 325/23/113/5 \\
    10 & 469 & 16,317(402) & 468 & 12,476 & 467(3) & 322/21/119/5 \\
    11 & 470 & 15,859(409) & 466 & 12,080 & 468(3) & 324/23/116/5 \\
    12 & 466 & 15,584(390) & --- & --- & 466(1) & 323/18/120/5 \\
    \bottomrule
  \end{tabular}
  \begin{flushleft}
    {\footnotesize Each column represents the following.\\
    \qquad $t$= month of the year 2019\\
    \qquad $|V_t|$= number of nodes\\
    \qquad $|E_t|$= number of edges (number of self-loops in parentheses)\\
    \qquad $|V_t\cap V_{t+1}|$= number of nodes common to successive months\\
    \qquad $|E_t\cap E_{t+1}|$= number of edges common to successive months\\
    \qquad GWCC= number of nodes in giant weakly connected component (number of components in parentheses)\\
    \qquad GSCC/IN/OUT/TE= number of nodes in giant strongly connected component/IN/OUT/tendrills}
  \end{flushleft}
  \label{tab:nw_basic}
\end{table}

We found that most of the users have self-loops, as shown in the
parentheses of column $|E_t|$, with the frequencies $f_{ii}$ and the
amounts $g_{ii}$ being highly correlated with the number of addresses
identified in the preceding \secref{sec:data_users} as naturally
expected. Because our main interest in this paper is the crypto flow
from one user to another, we remove all the self-loops in what follows.

Adjacent matrices with the strength of links given by the frequencies
$f_{ij}$ of $G_t$ for all $t$'s are illustrated in
Appendix~\ref{sec:adj}. One can see that the overall picture does not
change in time, but the illustration does not help to uncover the
nature of connectivity and flow.

To see the connectivity of network, namely how those regular users are
linked among them and also how they are located in the stream of cypto
flow, let us examine the property of connected components. First,
decompose $G_t$ into weakly connected components (WCC), i.e. connected
components when regarded as an undirected graph. We found that there
exists a giant WCC (GWCC) containing most of the users. See the
column GWCC of \tabref{tab:nw_basic}. There was only a small number
of disconnected components as shown in the same column.

Then, in order to identify the location of users contained in the
GWCC, we employed the well-known analysis of ``bow-tie'' structure
\cite{broder2000gsw}. In general, GWCC can be decomposed into the
following parts:
\begin{description}%
\setlength{\itemsep}{0pt}
\item[GSCC] Giant strongly connected component: the largest connected
  component when viewed as a directed graph. One or more directed
  paths exist for an arbitrary pair of firms in the component.
\item[IN] The nodes from which the GSCC is reached via at least one
  directed path.
\item[OUT] The nodes that are reachable from the GSCC via at least one
  directed path.
\item[TE] ``Tendrils''; the rest of the GWCC.
\end{description}
It follows that
\begin{equation}
  \label{eq:bowtie}
  \text{GWCC}=\text{GSCC}+\text{IN}+\text{OUT}+\text{TE}
\end{equation}
GSCC is the core of the crypto flow's circulation. The IN and OUT
parts are upstream and downstream of the flow respectively. The users
in the part of IN are playing a role of suppliers of crypto, while the
OUT users are considered to be consumers of crypto.

\tabref{tab:nw_basic} shows the bow-tie structure in the column of
GSCC/IN/OUT/TE. For example, in September, 470 users are located into
GSCC (325 users), IN (23), OUT (113), and TE (5). One can observe
that a large fraction of the users in the GWCC is located in the GSCC,
as one can easily interpret this fact in the way that those regular
users are circulating crypto globally. There are a less fraction of the
users in IN and OUT with asymmetry in the numbers.

\begin{figure}[tb]
  \centering
  \includegraphics[width=0.90\linewidth]{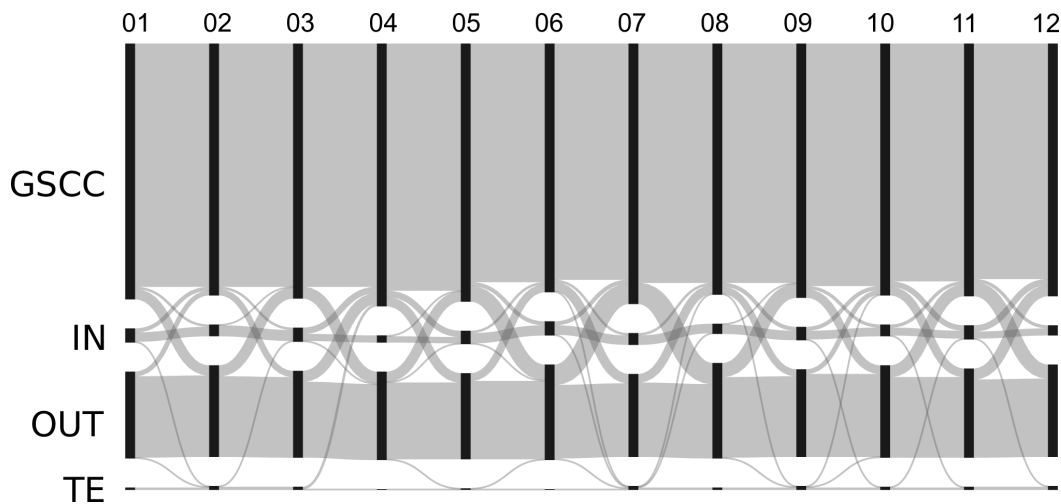}
  \caption{%
    Temporal change of bow-tie structure in an alluvial diagram.
    Monthly data of networks in the year 2019, January to December
    (from left to right). Vertical black segments in each month
    show the nodes of corresponding network grouped into
    GSCC (giant strongly connected component), IN, OUT, and
    TE (tendrills) in the bow-tie structure.
    Horizontal bands represent transitions among such groups
    from one month to its successive one.}
  \label{fig:bowtie_sankey}
\end{figure}

It would be interesting to see how the individual users are located in
the temporal change of the network. \figref{fig:bowtie_sankey}
depicts such a diagram of temporal change from one month to its
successive one in the whole year. One can see that the groups
of GSCC, IN, and OUT are very stable in each membership of users.
This fact means that those users appearing in successive months
are playing stable roles in the crypto flow's circulation and
the location of upstream and downstream. We remark that analysis
of bow-tie structure is based on the binary links, namely
either presence or absence among the nodes, but not on the
strength of links such as frequency and amount of crypto flow.
In the next section, we shall see how to quantify the location
of users by using the so-called Hodge decomposition.

\subsection{Hodge Decomposition}\label{sec:hodge}

Helmholtz-Hodge-Kodaira decomposition, or simply Hodge decomposition,
is a combinatorial method to decompose flow on a network into
circulation and gradient flow. Original idea dates back to the
Helmholtz theorem in vector analysis, which states that under
appropriate conditions any vector field can be uniquely represented by
the sum of an irrotational or rotation-free (curl-free) vector field
and a divergence-free (solenoidal) vector field. The theorem can be
generalized from Euclidean space to graph and other entity as shown by
Hodge, Kodaira and others. See \cite{jiang2008,jiang2011,johnson2013}
for readable exposition. The method has a wide range of applications
in the studies such as neural network \cite{miura2015shk},
economic networks \cite{kichikawa2018,fujiwara2021mfn}, and
also our previous work on Bitcoin and \cite{fujiwara2020hdb}.

We recapitulate the method briefly for the present manuscript to be
self-contained. Let $A_{ij}$ denote the adjacency matrix:
\begin{align}
  \label{eq:def_Aij}
  A_{ij} &=
  \begin{cases}
    1  & \text{if there is a link of transfer from user $i$ to $j$}, \\
    0  & \text{otherwise}.
  \end{cases}
\end{align}
We excluded all the self-loops, implying that $A_{ii}=0$.
Each link has a flow, denoted by $\tilde{F}_{ij}$, either of
the frequency, $f_{ij}$, or the amount, $g_{ij}$, of
the transfer from $i$ to $j$ (see \figref{fig:aggregate}).
Define
\begin{align}
  \label{eq:def_Bij}
  \tilde{F}_{ij} &=
  \begin{cases}
    f_{ij} \text{ or } g_{ij} & \text{if $A_{ij}=1$}, \\
    0  & \text{otherwise} .
  \end{cases}
\end{align}
Note that there can be a pair of users such that
$A_{ij}=A_{ji}=1$ and $\tilde{F}_{ij}, \tilde{F}_{ji}>0$.

Let us define a ``net flow'' $F_{ij}$ by
\begin{equation}
  \label{eq:def_Fij}
  F_{ij}=\tilde{F}_{ij}-\tilde{F}_{ji}
\end{equation}
and a ``net weight'' $w_{ij}$ by
\begin{equation}
  \label{eq:def_wij}
  w_{ij}=A_{ij}+A_{ji}.
\end{equation}
Note that $w_{ij}$ is symmetric, i.e., $w_{ij}= w_{ji}$,
and non-negative, i.e., $w_{ij}\geq 0$ for any pair of
$i$ and $j$\footnote{%
  It is remarked that \eqref{eq:def_wij} is simply a convention to consider
  the effect of mutual links between $i$ and $j$. One could multiply
  \eqref{eq:def_wij} by 0.5 or an arbitrary positive number, which
  does not change the result significantly for a large network.}.

\begin{figure}[tb]
  \centering
  \includegraphics[width=0.65\linewidth]{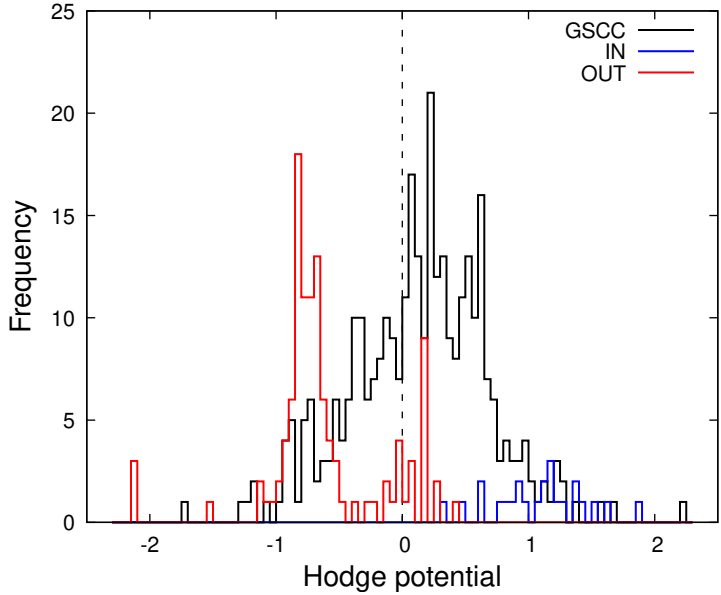}
  \caption{%
    Distributions for Hodge potentials of the users in
    GSCC (black), IN (blue), and OUT (red). The average of all the potential values
    is set to be zero (vertical dotted line). Data: 2019-09.}
  \label{fig:hodge_pot}
\end{figure}

Hodge decomposition is given by
\begin{equation}
  \label{eq:def_hodge}
  F_{ij}=F^{(\text{c})}_{ij}+F^{(\text{g})}_{ij}\ ,
\end{equation}
where the \textit{circular flow} $F^{(\text{c})}_{ij}$ satisfies
\begin{equation}
  \label{eq:def_Fcirc}
  \sum_j  F^{(\text{c})}_{ij}=0\ ,
\end{equation}
which implies that the circular flow is divergence-free.
The \textit{gradient flow} $F^{(\text{g})}_{ij}$ can be expressed as
\begin{equation}
  \label{eq:def_Fgrad}
  F^{(\text{g})}_{ij}=w_{ij}(\phi_i-\phi_j)\ .
\end{equation}
Thus the weight $w_{ij}$ serves to make the gradient flow possible
only where a link exists. We refer to the quantity $\phi_i$ as the
\textit{Hodge potential}. Large value of $\phi_i$ implies that the
user $i$ is in the upstream of the entire network, while small values
implies $i$ is in the downstream.

Combine \eqref{eq:def_hodge}, \eqref{eq:def_Fcirc}, and
\eqref{eq:def_Fgrad}, one can derive the following equation
to determine $\phi_i$.
\begin{equation}
  \label{eq:eqphi}
  \sum_j L_{ij} \phi_j = \sum_j F_{ij} \ ,
\end{equation}
for $i=1,\ldots,N$. Here, $L_{ij}$ is the so-called graph Laplacian and defined by
\begin{equation}
  \label{eq:def_Lij}
  L_{ij}=\delta_{ij} \sum_k w_{ik} - w_{ij} \ ,
\end{equation}
where $\delta_{ij}$ is the Kronecker delta.

It is easy to show that the matrix $L=(L_{ij})$ has only one zero mode
(eigenvector with zero eigenvalue).  The presence of this zero mode
simply corresponds to the arbitrariness in the origin of $\phi$.  All
the other eigenvalues are positive (see, e.g.,
\cite{fujiwara2020hdb}). Therefore, \eqref{eq:eqphi} can be solved for
the potentials by fixing the potentials' origin. We assume that the
average value of $\phi$ is zero.

\figref{fig:hodge_pot} depicts the distributions for Hodge potentials of
the users in GSCC, IN, and OUT. One can see that the entire set
of distributions is bimodal having two peaks at positive and negative
values, while there are a number of values around zero. Obviously,
they correspond to IN, OUT, and GSCC, each being located in
the upstream, downstream, and core of the entire crypto flow.
Moreover, there exists a correlation between the value of the Hodge
potential and the net amount of demand or supply of crypto by each
user.  See \cite{fujiwara2020hdb} for details, where we studied a
daily snapshot of the network including all the users, not only big
players. We claim that the same property holds also for the monthly
data restricted to big players of regular users.

\subsection{Non-negative Matrix Factorization}\label{sec:nmf}

It would be a natural question whether there are distinctive
ingredients of flows in the crypto flow or not. The analysis of
bow-tie structure is based merely on the binary relationship of links,
so does not give such information, because the crypto flows from
upstream to downstream with circulation in the giant strongly
connected component that occupies a large fraction of the entire
network. In other words, are there any ``principal components'' that
constitute the entire flow in a decomposition? In order to find such
principal components or latent factors in the transfer of crypto among
big players, we shall apply \textit{non-negative matrix factorization}
(NMF) to the strength of links, namely the matrix of the frequencies
and amounts of transfer. We recapitulate the method here. See
\cite{LeeSeung1999,berry2007aaa,gaujoux2010frp} and references
therein for introduction.

Let $X$ be an $N\times M$ non-negative matrix, in general, to start
with; that is, its elements are all non-negative, denoted as
$X\geq 0$. NMF gives an approximation of $X$ by a product of two
matrices:
\begin{equation}
  \label{eq:def_nmf}
  X\approx S D\ ,
\end{equation}
where $S,D$ are $N\times K$ and $K\times M$ non-negative matrices,
$S,D\geq 0$, respectively\footnote{%
  The decomposition is not unique due to trivial degrees of freedom.
  One is permutation, $SD=S\pi\pi^{-1}D$, where $\pi$ is a permutation
  matrix simply exchanging indices. Another is scale transformation,
  $SD=S\sigma\sigma^{-1}D$, where $\sigma$ is a diagonal matrix
  with all elements positive. We shall see that these degrees are fixed
  after appropriate normalization and ordering.}%
. In practice, one expects that $K$ is much smaller than $N$ and $M$
so that the factorization gives a compact representation of $X$.
We shall assume that $N=M$ for our application of crypto flow
among $N$ users in what follows.

Explicitly in components, \eqref{eq:def_nmf} reads
\begin{equation}
  \label{eq:def_nmf2}
  X_{sd}\approx\sum_{k=1}^K S_{sk} D_{kd}\ ,
\end{equation}
where the indices $s$ and $d$ represent source and destination
($s,d=1,\ldots,N$) respectively, and $X_{sd}$ is the strength of
crypto flow, quantified by frequency $f_{sd}$, amount $g_{sd}$, or
similar variables, from $s$ to $d$ in a certain period of time.
We choose
\begin{equation}
  \label{eq:Xandf}
  X_{sd}=f_{sd}\ ,
\end{equation}
in this paper. See \figref{fig:adj} in Appendix~\ref{sec:adj}
for the illustration of $X_{sd}$.
We would expect that $K\ll N$, because of the sparsity of $X$.
How to determine $K$ is discussed later.

The approximation in \eqref{eq:def_nmf} is actually given
by the following optimization:
\begin{equation}
  \label{eq:opt_nmf}
  \min_{S,D\geq 0} F(X, S D)\ ,
\end{equation}
where the function $F(\cdot,\cdot)$ is the so-called
Kullback-Leibler (KL) divergence defined by
\begin{equation}
  \label{eq:def_KL}
  F(A,B)=\text{KL}(A\Vert B)\equiv
  \sum_{i,j}\left(A_{ij}\log\frac{A_{ij}}{B_{ij}}-A_{ij}+B_{ij}\right)\ .
\end{equation}
Note that $F(A,B)=0$ if and only if $A=B$.
The reason why we choose the particular function of 
\eqref{eq:def_KL} will be clarified later\footnote{%
  See \secref{sec:nmf_prob} for a probabilistic interpretation of
  choosing the KL divergence. Another functional form, often used,
  is the so-called Frobenius norm:
  $F(A,B)=(1/2)\sum_{i,j}(A_{ij}-B_{ij})^2$, which leads to
  a different probabilistic model.}%
. Technically, one can solve \eqref{eq:opt_nmf} iteratively with
the initialization of $S,D$ using non-negative double singular value
decomposition (see the review \cite{berry2007aaa} and references
therein). Although the iterative algorithm yields local minima,
our numerical solutions under different random seeds gave
essentially the same decomposition.

To understand the meaning of the decomposition, let us consider
how a source distributes flow to different destinations.
For an arbitrary source $s$, \eqref{eq:def_nmf2} can be written as
\begin{equation}
  \label{eq:nmf_Xs}
  \bm{X}_s\approx\sum_{k=1}^K S_{sk}\bm{D}_k\ ,
\end{equation}
where $\bm{X}_s$ is the vector of $s$-th row of $X$,
and $\bm{D}_k$ is the vector of $k$-th row of $D$.
Equation \eqref{eq:nmf_Xs} means that the flow from the source $s$
can be expanded in terms of ``basis'' vectors,
$\bm{D}_k$ ($k=1,\ldots,K$). The components
$\left(\bm{D_k}\right)_d=D_{kd}$ represent how
\textit{destinations} are distributed among users in
the $k$-th NMF component. It is convenient to
normalize $\bm{D}_k$ by L1-norm, that is,
by defining
\begin{equation}
  \label{eq:def_Dnorm}
  \widetilde{D}_{kd}\equiv\frac{D_{kd}}{D_k}
  \quad\text{where}\quad
  D_k\equiv\sum_{d} D_{kd}\ ,
\end{equation}
so that one has
\begin{equation}
  \label{eq:Dnorm}
  \sum_d \widetilde{D}_{kd}=1\ ,
\end{equation}
for all $k$. With respect to this normalized basis vectors,
the expansion in \eqref{eq:nmf_Xs} is rewritten as
\begin{equation}
  \label{eq:nmf_Xs2}
  \bm{X}_s\approx
  \sum_{k=1}^K (S_{sk}D_k)\bm{\widetilde{D}}_k
  \quad\text{where}\quad
  (\bm{\widetilde{D}}_k)_d=\widetilde{D}_{kd}\ ,
\end{equation}
Thus the outgoing flow from the source $s$ is approximately expressed by
a linear combination of $K$ normalized basis vectors $\bm{\widetilde{D}}_k$
with coefficients given by $S_{sk}D_k$.

Similarly, consider how a destination $d$ collects flow
from different sources. For an arbitrary destination $d$,
\eqref{eq:def_nmf2} reads
\begin{equation}
  \label{eq:nmf_Xd}
  \bm{X}_d\approx\sum_{k=1}^K D_{kd}\bm{S}_k\ ,
\end{equation}
where $\bm{X}_d$ is the vector of $d$-th column of $X$,
and $\bm{S}_k$ is the vector of $k$-th column of $S$.
The components of $(\bm{S}_k)_s=S_{sk}$ represent how
\textit{sources} are distributed among users in
this $k$-th NMF component. Define
\begin{equation}
  \label{eq:def_Snorm}
  \widetilde{S}_{sk}\equiv\frac{S_{sk}}{S_k}
  \quad\text{where}\quad
  S_k\equiv\sum_{s} S_{sk}\ ,
\end{equation}
and one has
\begin{equation}
  \label{eq:Snorm}
  \sum_s \widetilde{S}_{sk}=1\ ,
\end{equation}
for all $k$. Then \eqref{eq:nmf_Xd} is rewritten as
\begin{equation}
  \label{eq:nmf_Xd2}
  \bm{X}_d\approx
  \sum_{k=1}^K (D_{kd}S_{k})\bm{\widetilde{S}}_k
  \quad\text{where}\quad
  (\bm{\widetilde{S}}_k)_s=\widetilde{S}_{sk}\ .
\end{equation}
Thus the incoming flow to the destination $d$ is approximately expressed by
a linear combination of $K$ normalized basis vectors $\bm{\widetilde{S}}_k$
with coefficients given by $D_{kd}S_k$.

How can one determine $K$? Obviously, the larger $K$ is, the better
the approximation \eqref{eq:def_nmf} is, but with less parsimonious
representation of the data. In the next section, let us make a detour
to examine this issue from a different perspective.

\subsection{NMF as a probabilistic model}\label{sec:nmf_prob}

We can interpret the NMF as a probabilistic model. Denote the
right-hand of \eqref{eq:def_nmf2} by
\begin{equation}
  \label{eq:def_xi}
  \xi_{sd}\equiv\sum_k S_{sk}D_{kd}\ ,
\end{equation}
which are regarded as parameters to be estimated from the data $X$
such that $X_{sd}$ is assumed to be a random number chosen from
a Poisson distribution with the parameter $\xi_{sd}$ as
\begin{equation}
  \label{eq:def_poi}
  P(x\,|\,\xi)=e^{-\xi}\,\frac{\xi^x}{x!}\ .
\end{equation}
It is easy to see that the log likelihood function
$L(\xi_{sd})\equiv\log P(X_{sd}\,|\,\xi_{sd})$ takes
the maximum value at $\xi_{sd}=X_{sd}$.
Then one can introduce a quantity to measure
how much the estimation of the parameters is good, that is
\begin{equation}
  \label{eq:KL_poi}
  \sum_{s,d}\left(L(X_{sd})-L(\xi_{sd})\right)=
  \sum_{s,d}\left(X_{sd}\log\frac{X_{sd}}{\xi_{sd}}-X_{sd}+\xi_{sd}\right)\ ,
\end{equation}
to be \textit{minimized}. One can see that this quantity is equivalent
to the KL divergence in \eqref{eq:def_KL}\footnote{%
  We became aware that this argument is known in the literature.
  If one assumes Gaussian instead of Poisson, one would have
  Frobenius norm for KL divergence in \eqref{eq:def_KL}.
  See \cite{kameoka2015nmf}. One of the present authors (YF) learned a
  hint on the argument from Itsuki Noda in his application of NMF to
  transportation data \cite{noda2019nmf}.}.

To express the entire framework in probabilistic terms more explicitly,
let us normalize the data $X$ in \eqref{eq:Xandf} by
\begin{equation}
  \label{eq:def_Xnorm}
  \widetilde{X}_{sd}=\frac{X_{sd}}{\sum_{s',d'} X_{s'd'}}\ .
\end{equation}
Then let us rewrite \eqref{eq:def_nmf2} as
\begin{equation}
  \label{eq:def_nmf3}
  \widetilde{X}_{sd}\approx\sum_k r_k \widetilde{S}_{sk} \widetilde{D}_{kd}\ ,
\end{equation}
where $\widetilde{D}_{kd}$ and $\widetilde{S}_{sk}$ were given by
\eqref{eq:def_Dnorm} and \eqref{eq:def_Snorm} respectively, and
\begin{equation}
  \label{eq:def_rk}
  r_k\equiv\frac{S_k D_k}{\sum_{k'} S_{k'} D_{k'}}\ ,
\end{equation}
which satisfies that $\sum_k r_k=1$.
Let us denote the right-hand side of \eqref{eq:def_nmf3} by
\begin{equation}
  \label{eq:def_psd}
  p_{sd}\equiv\sum_k r_k \widetilde{S}_{sk} \widetilde{D}_{kd}\ ,
\end{equation}
which satisfies that $\sum_{s,d}p_{sd}=1$.
We remark that the normalized weight $r_k$ defined by
\eqref{eq:def_rk} gives the information of relative importance
of the $k$-th NMF component in the expansion with
normalized basis vectors in \eqref{eq:def_psd}.
One can determine the ordering of NMF components uniquely
according to the magnitudes of $r_k$.

Suppose that there are $N_f$ transfers in total during a period of
time. For each pair of source and destination, $s$ and $d$, generate a
transfer $s\rightarrow d$ with the probability given by $p_{sd}$,
being independently of other pairs. Under the assumption of
a small probability of $p_{sd}$ and a large number of $N_f$,
$X_{sd}$ follows a Poisson distribution with the parameter,
$\xi_{sd}=N_f p_{sd}$.

It turns out that the decomposition in \eqref{eq:def_xi}, or
equivalently \eqref{eq:def_psd}, has an interesting connection with
machine learning. In natural language processing, it is often
necessary to extract \textit{topics} among documents comprising of
words or terms. In a situation of unsupervised learning, the task is
to infer topics as hidden or latent variables, which can explain a
collection of documents, each being an unordered set of
terms. Probabilistic latent semantic analysis (PLSA) is a probabilistic
model for doing such a task \cite{hofmann2001plsa}.
Suppose that there are $N$ documents and
$M$ terms. Then the occurrence of terms can be expressed by a
document-term matrix $X$ with size $N\times M$, each element of which
is the frequency of occurrence of a term in a document. Topics are
latent variables to explain the data $X$. A topic is actually a
probability distribution for the occurrence of terms with different
probabilities. A document can have a mixture of topics. An example is
a document on ``influence of hosting Olympics to economy'' with a
mixture of topics on sports and economy.

One of the widely used model of PLSA is latent Dirichlet allocation
(LDA). See Appendix~\ref{sec:lda} and references therein. For our
purpose, it suffices to understand how terms are generated at
locations of documents in a probabilistic way. The probability that a
term is chosen at a location in a document is given by the sum of $K$
factors, each of which is the product of two probabilities; the
probability that a topic is selected in the document and the one that
the term is chosen under the selected topic. See the equation of
\eqref{eq:lda_nmf} in Appendix~\ref{sec:lda}. One can immediately see
that \eqref{eq:lda_nmf} is essentially the same as
\eqref{eq:def_xi}, or equivalently \eqref{eq:def_psd}.

Thus the matrix decomposition of NMF can be put in the framework of
probabilistic model of PLSA and LDA. As a bonus, one can adopt the
method of estimating the number of topics to our problem of determining
the number of NMF components, denoted by $K$ in both cases.
Interested readers are guided to look at the literature
\cite{griffiths2004fst,cao2009dbm,arun2010fnn,deveaud2014ael} and
others given at the end of Appendix~\ref{sec:lda}. Let us take a look
at our results in the next section finishing the detour of this section.

\subsection{Result of NMF for Crypto Flow}\label{sec:nmf_res}

\begin{figure}[tb]
  \centering
  \subfigure[Comparison of three methods \cite{cao2009dbm,arun2010fnn,deveaud2014ael}]{
    \includegraphics[width=0.45\linewidth]{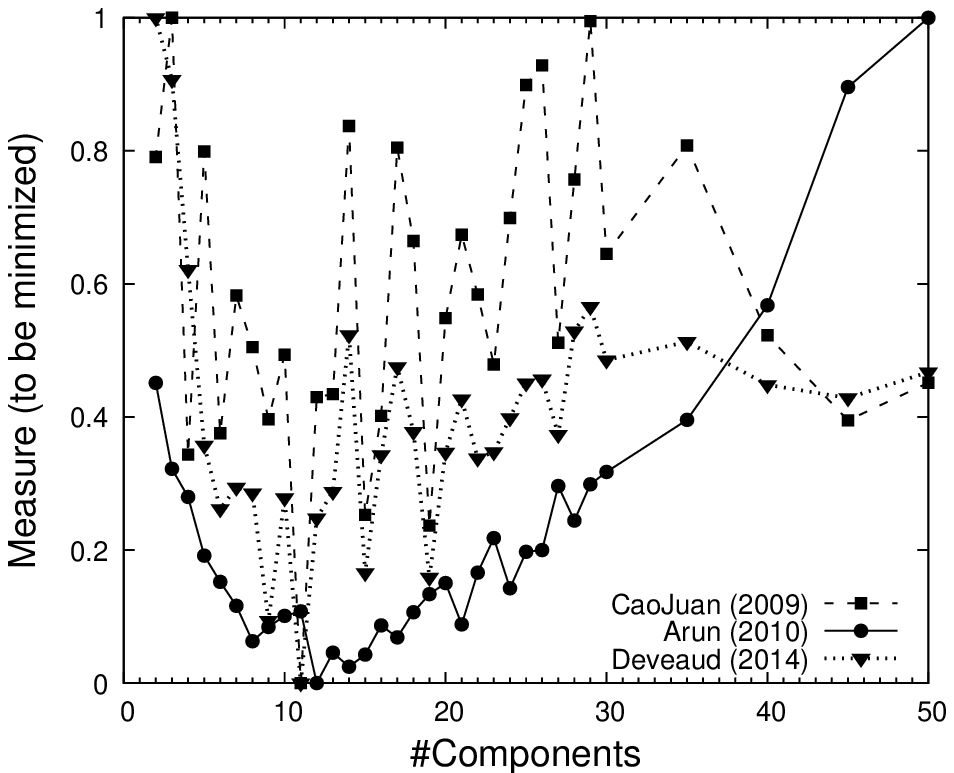}
    \label{fig:detK1909a}}
  \hspace{2em}
  \subfigure[Monte Carlo simulations by the method \cite{arun2010fnn}]{
    \includegraphics[width=0.45\linewidth]{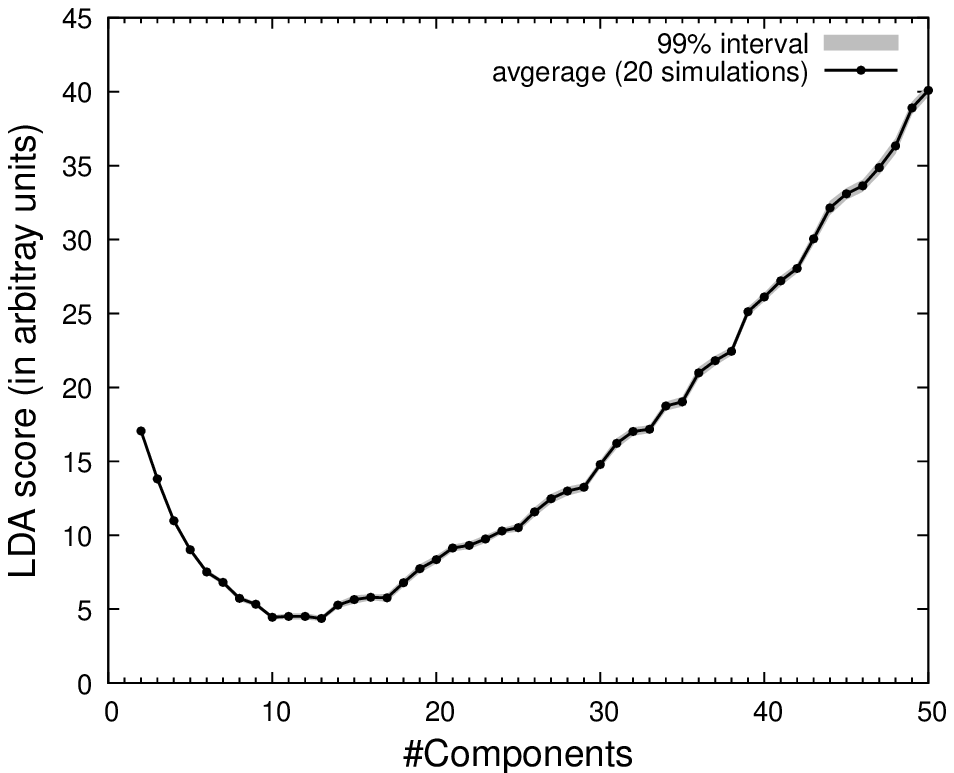}
    \label{fig:detK1909b}}
  \caption{%
    Determining $K$, the number of NMF components, by using the
    concept of coherence in LDA. %
    (a)~Different methods \cite{cao2009dbm,arun2010fnn,deveaud2014ael}
    are compared. Each measure of coherence is drawn in the vertical
    axis so that it is to be minimized to find the optimal number of
    components. Maximum and minimum values in the region of $K$
    are scaled to 1 and 0 respectively to make the comparison easier.
    One can see that $K=11\sim 13$ are optimal. %
    (b)~Monte Carlo simulations by the method \cite{arun2010fnn} with
    20 runs for each $K$. Averages (points) and 99\% level (gray band, narrow)
    calculated from standard errors are drawn. We conclude that
    $K=13$ is optimal from this result (b).}
  \label{fig:detK1909}
\end{figure}

We first show a few results for a snapshot of September in the year 2019
(denoted as 2019-09), in order to verify if the idea
in the preceding section works to determine the number of NMF
components. \figref{fig:detK1909} shows the measures
of coherence by employing three different methods of LDA
\cite{cao2009dbm,arun2010fnn,deveaud2014ael}. The methods
give mostly the same optimal values of $K$,
namely $K=11\sim 13$, consistently as shown in
\figref{fig:detK1909}~(a).
We found that the measure given in \cite{arun2010fnn}
is relatively stable and potentially useful to determine
a specific value of $K$. So we performed Monte Carlo simulations
in \figref{fig:detK1909}~(b), and were able to determine
the optimal value as $K=13$. For this data, $X$ has
the dimension of $N=470$, so we conclude that one can have
a small number of NMF components that can explain the entire flow among
those regular users.

In Appendix~\ref{sec:optnum}, we summarize the same result for the
data in all the other months of the year 2019. We found that the
optimal number $K$ is quite small in the range more than 10 and less
than 20, much smaller than the number of users, $N\sim 500$ (see
\tabref{tab:nw_basic}). Additionally, $K$ is relatively stable
irrespectively of the temporal change.
See \tabref{tab:optnum} and \figref{fig:fig_ldat_all}.

\begin{figure}[tb]
  \centering
  \includegraphics[width=0.98\linewidth]{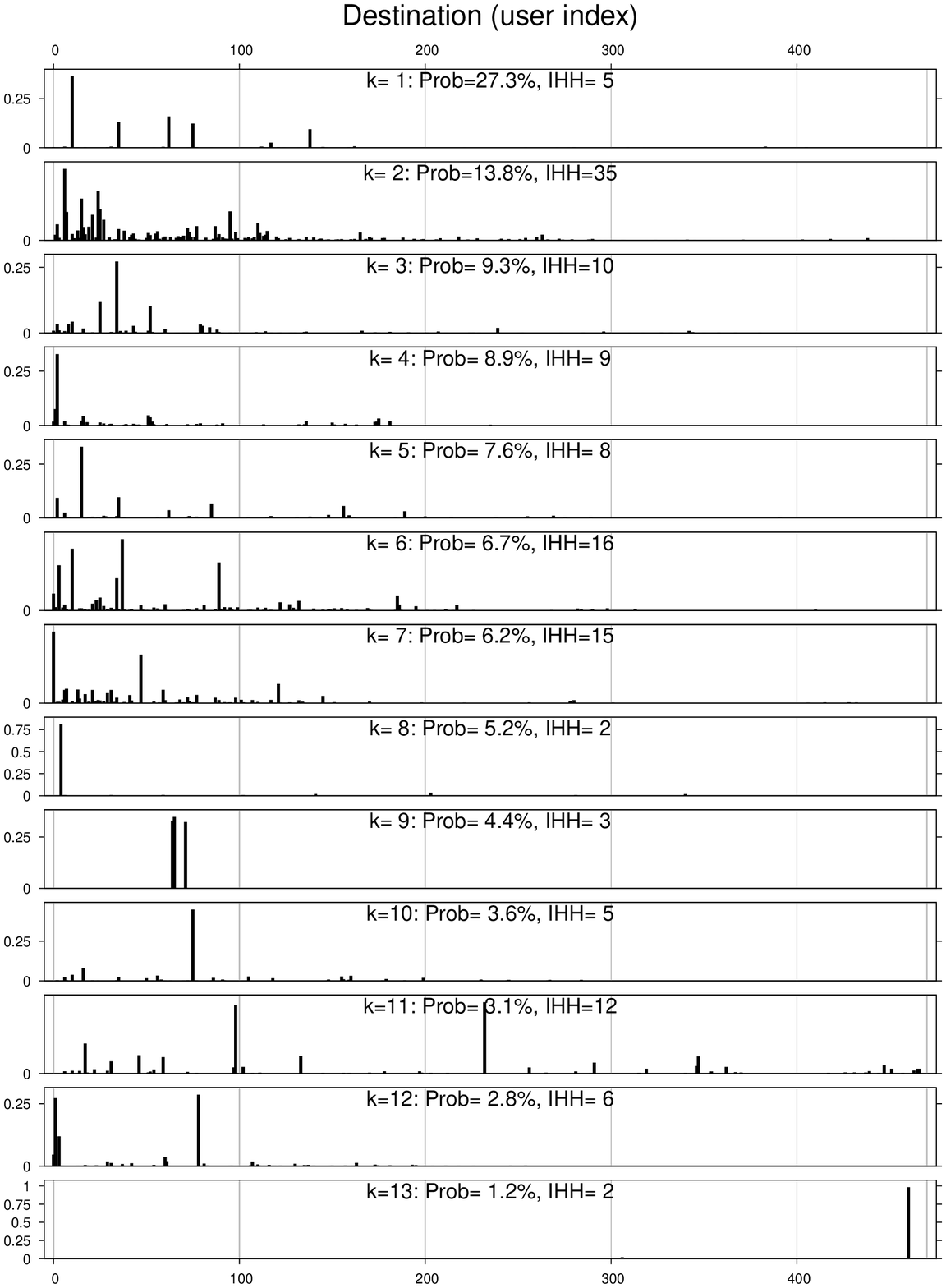}
  \caption{%
    NMF components $\bm{\widetilde{D}}_k$ ($k=1,\ldots,13$) from top to bottom.
    Each plot $(\bm{\widetilde{D}}_k)_d=D_{kd}$ shows how destinations are distributed
    among users $d$ in the $k$-th NMF component. Note that $\bm{\widetilde{D}}_k$
    is normalized, i.e. $\sum_d D_{kd}=1$ for all $k$.
    See \eqref{eq:def_Dnorm} and \eqref{eq:Dnorm} in the main text.
    Also shown in each plot are the probability of the component, denoted by ``Prob'',
    and the inverse Herfindahl-Hirschman index ``IHH'' representing
    the effective number of dominant users in the component. Data: 2019-09.}
  \label{fig:Hnorm_2019-09}
\end{figure}

\begin{figure}[tb]
  \centering
  \includegraphics[width=0.98\linewidth]{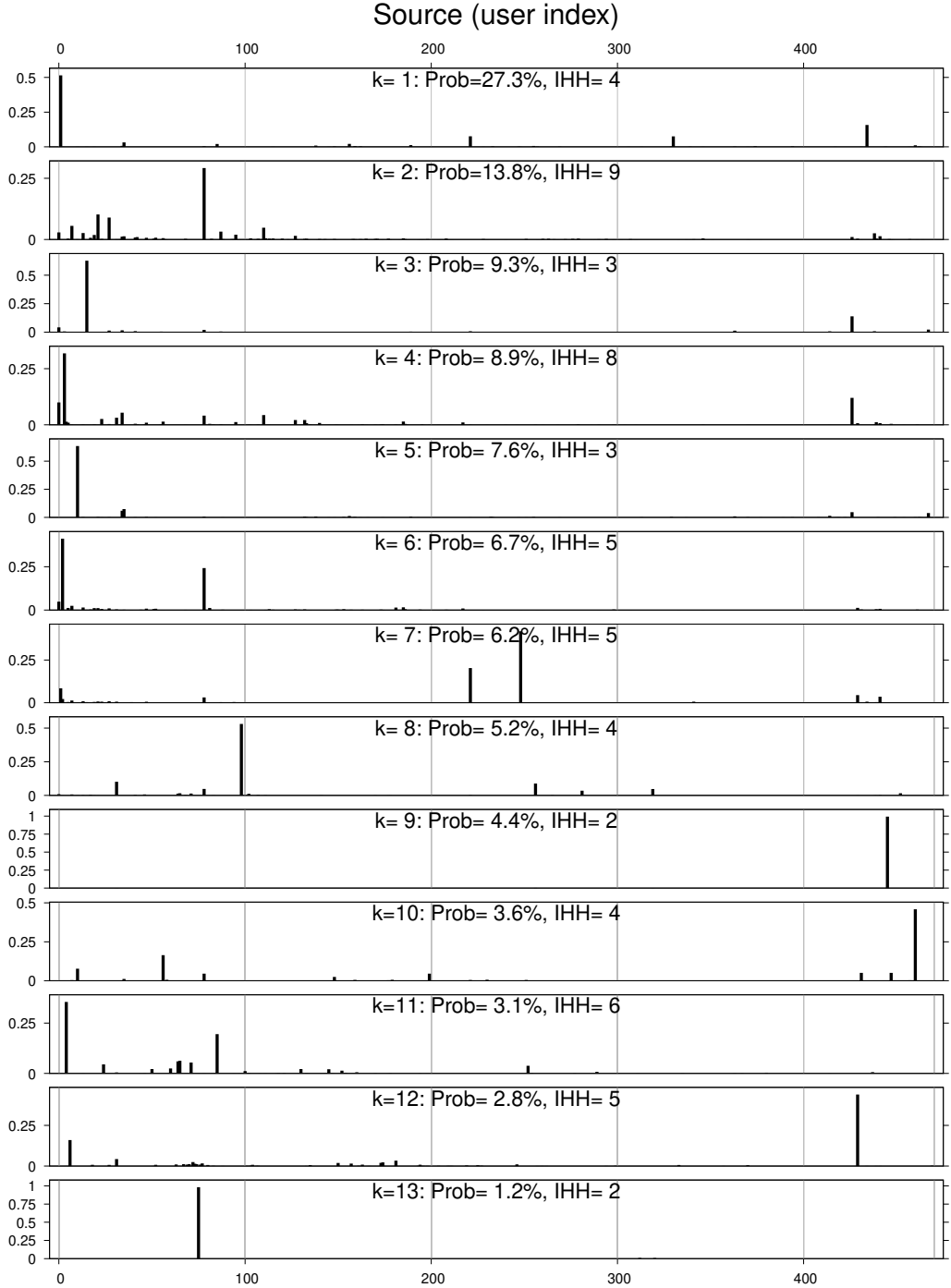}
  \caption{%
    NMF components $\bm{\widetilde{S}}_k$ ($k=1,\ldots,13$) from top to bottom.
    Each plot $(\bm{\widetilde{S}}_k)_s=S_{sk}$ shows how sources are distributed
    among users $s$ in the $k$-th NMF component. Note that $\bm{\widetilde{S}}_k$
    is normalized, i.e. $\sum_s S_{sk}=1$ for all $k$.
    See \eqref{eq:def_Snorm} and \eqref{eq:Snorm} in the main text.
    See the caption of \figref{fig:Hnorm_2019-09} for ``Prob'' and ``IHH'' in each plot.
    Data: 2019-09.}
  \label{fig:Wnorm_2019-09}
\end{figure}

\begin{figure}[tb]
  \centering
  \includegraphics[width=0.85\linewidth]{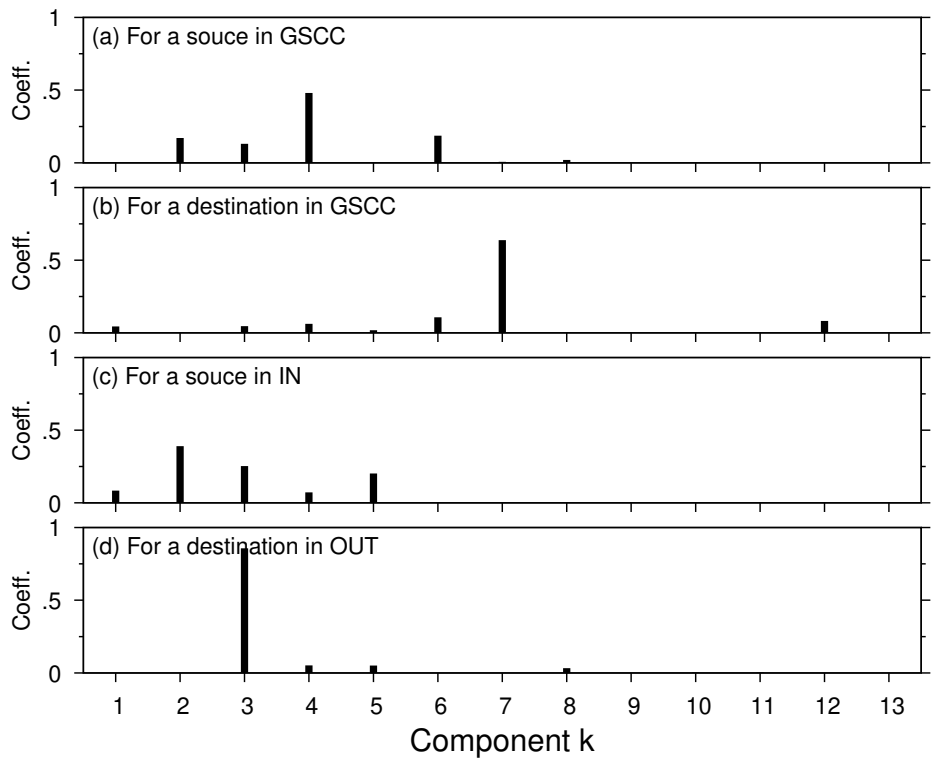}
  \caption{%
    Coefficients, with which crypto flow from or to a selected user
    is expanded with respect to the NMF components. The selected users
    are \texttt{0000000000} (a,b) included in the GSCC of bow-tie
    structure, \texttt{0000006178} (c) in the IN, and
    \texttt{0000000012} (d) in the OUT. The user \texttt{0000000000} can be
    source (a) and destination (b). The user of (c) is a source,
    and the user of (d) is a destination. The expansion is given by
    \eqref{eq:nmf_Xs2} for the selected source $s$, and by
    \eqref{eq:nmf_Xd2} for the selected destination $d$.
    Data: 2019-09.}
  \label{fig:nmf_coeff}
\end{figure}

\begin{figure}[tb]
  \centering
  \includegraphics[width=0.85\linewidth]{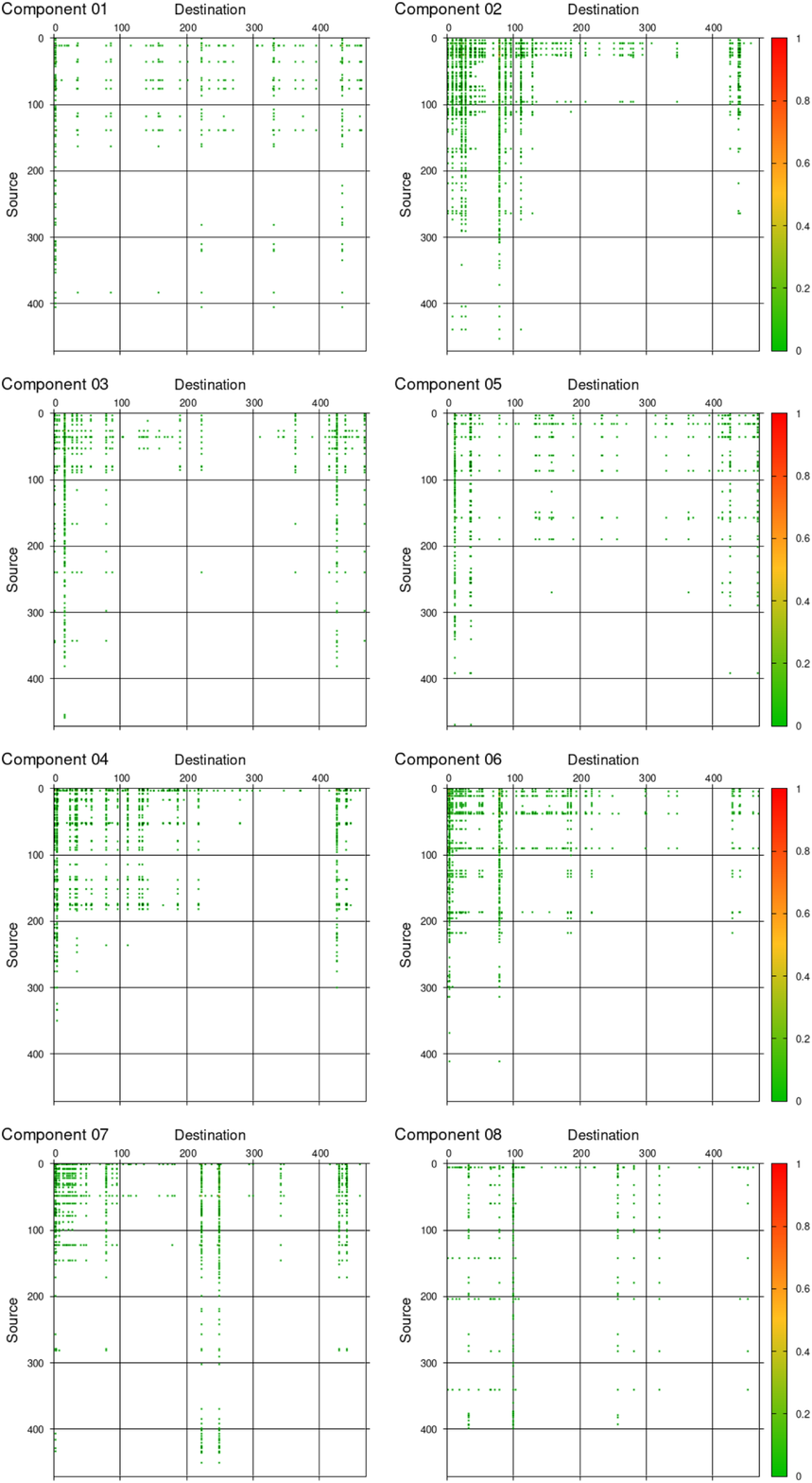}
  \caption{%
    NMF components $k=1,\ldots,8$ as matrices defined by
    $\widetilde{S}_{sk} \widetilde{D}_{kd}$ in
    \eqref{eq:def_psd}. Data: 2019-09.}
  \label{fig:comp_mat_a}
\end{figure}

\begin{figure}[tb]
  \centering
  \includegraphics[width=0.85\linewidth]{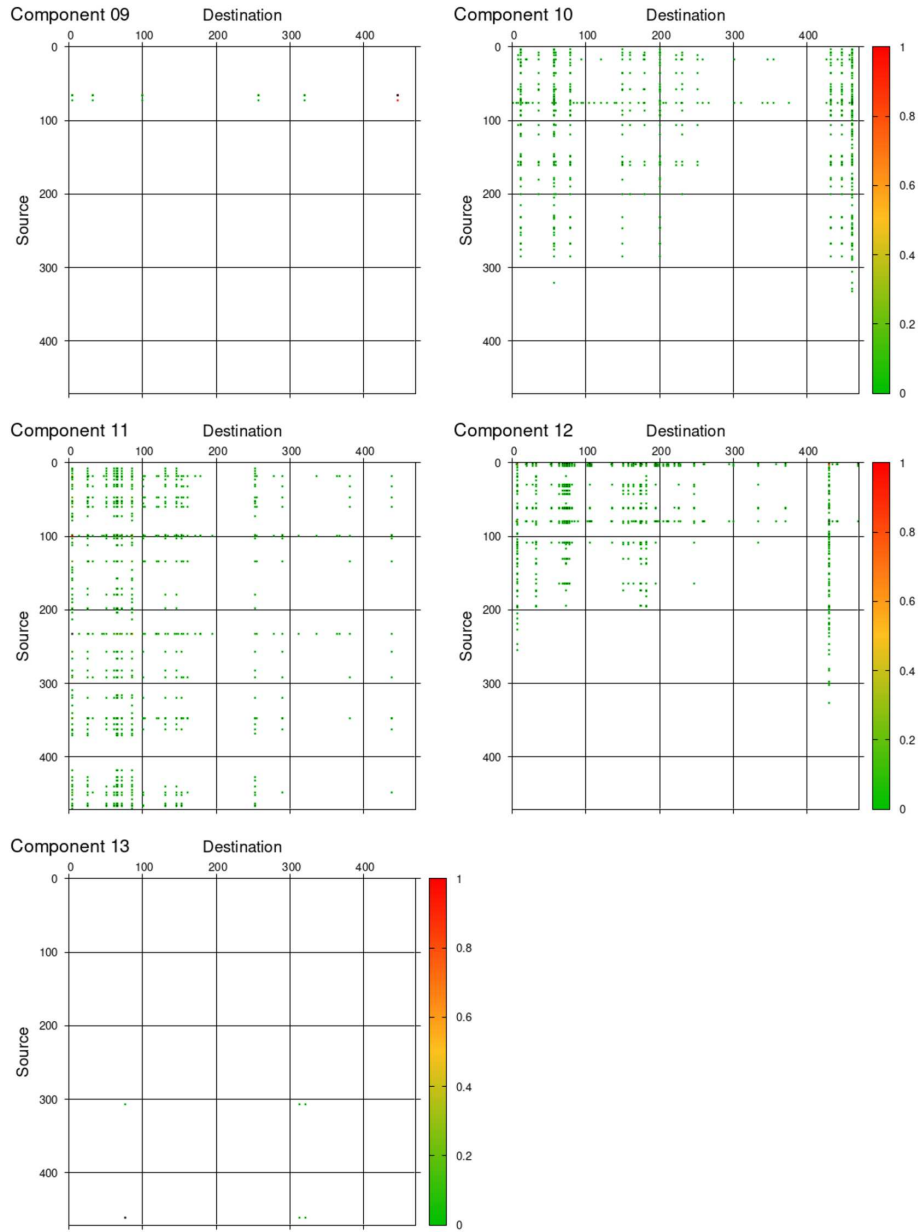}
  \caption{%
    (Continued) NMF components $k=9,\ldots,13$ as matrices defined by
    $\widetilde{S}_{sk} \widetilde{D}_{kd}$ in
    \eqref{eq:def_psd}. Data: 2019-09.}
  \label{fig:comp_mat_b}
\end{figure}

\begin{figure}[tb]
  \centering
  \subfigure[Cosine similarity of NMF basis vectors $\bm{D}_k$]{
    \includegraphics[width=0.70\linewidth]{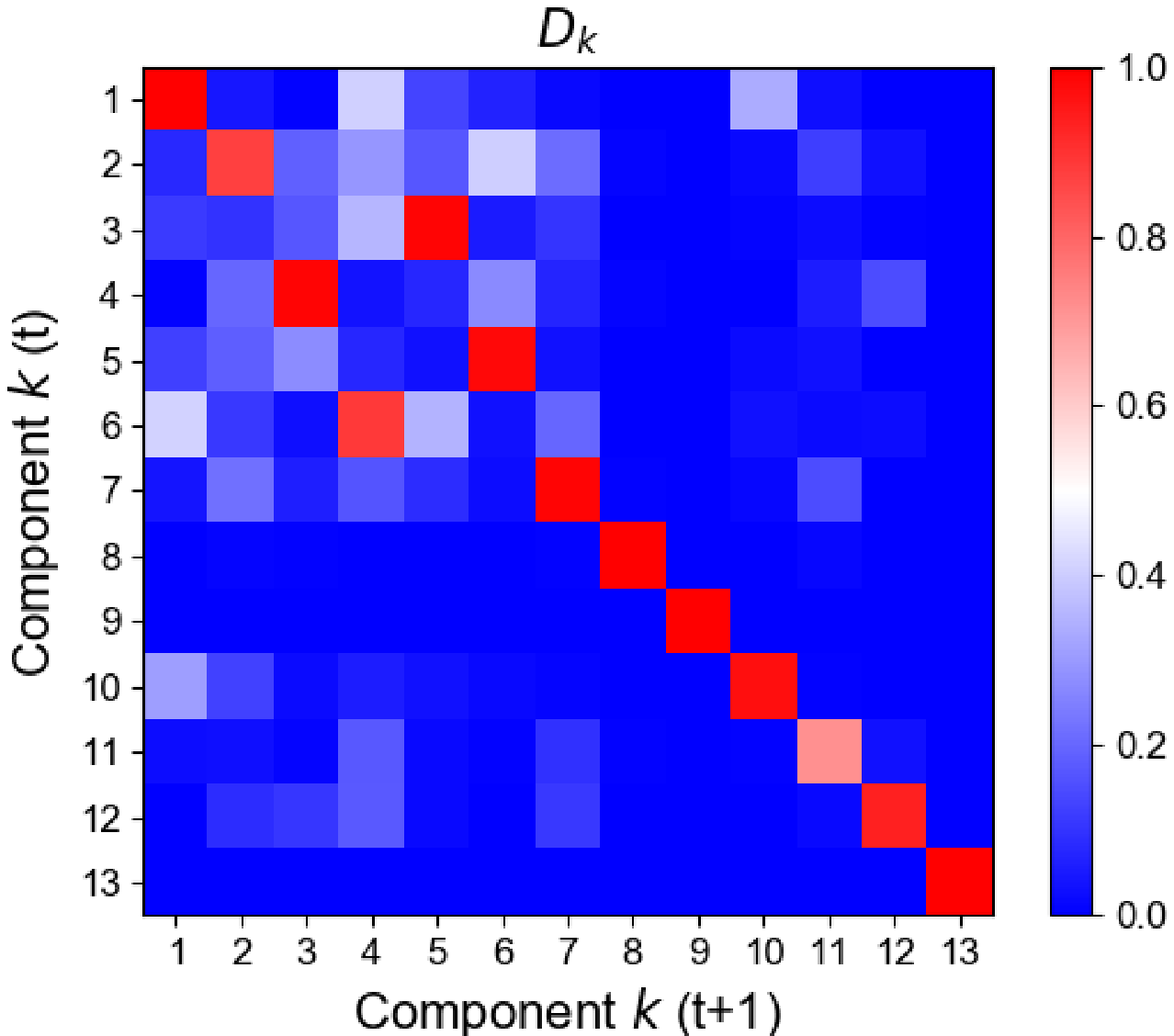}}
  \\
  \subfigure[Cosine similarity of NMF basis vectors $\bm{S}_k$]{
    \includegraphics[width=0.70\linewidth]{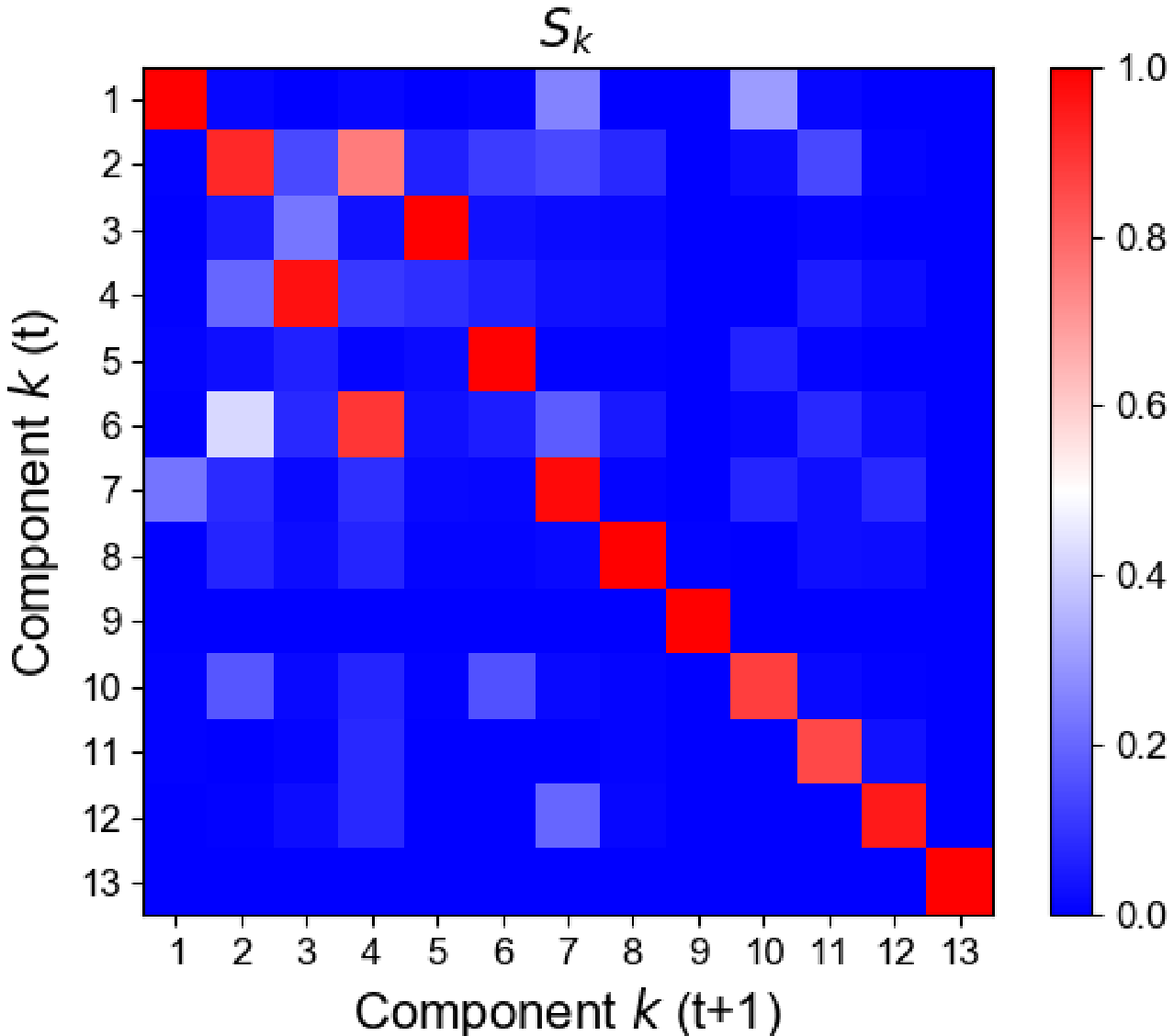}}
  \caption{%
    Temporal change of the NMF components from one month $t$ to its
    successive month $t+1$. %
    (a)~Cosine similarities calculated for all the pairs of $\bm{D}_k$
    ($k=1,\ldots,13$) between $t$ (vertical) and $t+1$. %
    (b)~The same for all the pairs of $\bm{S}_k$. %
    For the vertical and horizontal axes in (a) and (b), the order of
    indices along vertical and horizontal axes corresponds to the
    descending order of the probability $r_k$ in \eqref{eq:def_rk}.
    Data: 2019-09 and 2019-10.}
  \label{fig:comp_cf}
\end{figure}

Let us examine each NMF components obtained with the optimal value of
$K$. \figref{fig:Hnorm_2019-09} and \figref{fig:Wnorm_2019-09} show
the NMF components in terms of the basis vectors,
$\bm{\widetilde{D}}_k$ and $\bm{\widetilde{S}}_k$, respectively for
$k=1,\ldots,K$. In \figref{fig:Hnorm_2019-09}, each plot shows the
vector components of $(\bm{\widetilde{D}}_k)_d=D_{kd}$ meaning how
destinations are distributed among users $d$ in the $k$-th NMF
component. Similarly in \figref{fig:Wnorm_2019-09}, each plot shows
the vector components of $(\bm{\widetilde{S}}_k)_s=S_{sk}$ meaning how
sources are distributed among users $s$ in the $k$-th NMF component.
See \eqref{eq:def_Dnorm} and \eqref{eq:def_Snorm}, and also note the
normalization therein. Note that in each of \figref{fig:Hnorm_2019-09}
and \figref{fig:Wnorm_2019-09}, the plots are ordered (from top to
bottom) in the descending order of the probability $r_k$ given in
\eqref{eq:def_rk}.

One can immediately notice from the figures that the components of
these basis vectors are concentrated on a limited number of users, but
are not distributed among many users. To quantify the effective number
of the concentration, let us use the inverse Herfindahl-Hirschman
index, abbreviated as IHH, which is defined as follows. Consider
``shares'' $x_i\geq 0$ among $i=1,\ldots,N$ things with the sum
equal to 1, i.e. $\sum_i x_i=1$. The IHH is defined by
\begin{equation}
  \label{eq:def_IHH}
  \text{IHH}\equiv\left(\sum_{i=1}^N x_i^2\right)^{-1}\ .
\end{equation}
When the shares are equal, $x_i=1/N$ for all $i$, then
$\text{IHH}=N$. On the other hand, when there is the strongest
concentration, namely, $x_i=1$ for a particular $i$ and
$x_i=0$ otherwise, then $\text{IHH}=1$. So IHH can give
an estimate of the effective number of large shares\footnote{%
  It is remarked that the paper \cite{akky2021pc} in this volume
  also applied but modifies Herfindahl-Hirschman index
  in an interesting way to characterize the frequencies of
  transactions in XRP.}. %
The idea can be applied to the basis vectors, because
the vectors are normalized in the same way as shares.
In \figref{fig:Hnorm_2019-09} and \figref{fig:Wnorm_2019-09},
we displayed all the calculated IHH's. One can see that
the IHH's are quite small ranging from a few to a dozen or so,
compared with the total number of users $N=470$
for the data 2019-09 (see \tabref{tab:nw_basic}).

How can we use the NMF components to understand the crypto flow?
Choose a particular user $s$ as a source $s$. The flow from $s$ was approximately
expressed by a linear combination of $K$ normalized basis vectors
$\bm{\widetilde{D}}_k$, each depicted in \figref{fig:Hnorm_2019-09},
with the coefficients given in \eqref{eq:nmf_Xs2},
i.e. $S_{sk}D_k$. The coefficients represent the strength of the
decomposed flow from the source $s$. A similar argument holds
by choosing a particular user $d$ as a destination. The flow to
$d$ was expressed by a linear combination in \eqref{eq:nmf_Xd2}
with the coefficients, $D_{kd}S_k$.

For example, consider the user with ID \texttt{0000000000} that is located
in the GSCC, as a source $s$. \figref{fig:nmf_coeff} shows the
coefficients corresponding to $K$ components; see (a). One can see
that the coefficients are non-zero at only four components. Such a
sparseness tells that the flow from this user can be expressed with a
few components. And the corresponding components have non-zero values
at a small number (recall the IHH's) of the vector components,
$D_{kd}$, as shown in \figref{fig:Hnorm_2019-09}, implying that the
users corresponding to these non-zero components constitute
a cluster for the outgoing flow from the source $s$.

The same user \texttt{0000000000} can be regarded as a destination
$d$ in the GSCC. \figref{fig:nmf_coeff}~(b) shows the coefficients,
again non-zeros at only one or two components. Together with
\figref{fig:Wnorm_2019-09}, one can find another cluster composed
of a small number of users for the incoming flow to the $d$.
Similar arguments hold for the users \texttt{0000006178} and
\texttt{0000000012}, respectively located in the IN and OUT.
See \figref{fig:nmf_coeff}~(c) and (d). In this way, one can
find clusters for either of or both of the outgoing and incoming
flows of each user.

Each NMF component can be represented by a matrix, because the
non-negative matrix of $X_{sd}$ or its probabilistic counter part
$p_{sd}$ can be expressed by \eqref{eq:nmf_Xs} or \eqref{eq:def_psd}.
It is possible to depict each component $k$ by the matrix of
$\widetilde{S}_{sk} \widetilde{D}_{kd}$ in the normalized way.
\figref{fig:comp_mat_a} and \figref{fig:comp_mat_b} illustrate
such matrices for the data of 2019-09. This should be compared
with $X_{sd}$ in Appendix~\ref{sec:adj}. One can see that
the NMF components provide sparse matrices.

Finally, we find that the NMF components are relatively stable
in the temporal change of network. \figref{fig:comp_cf}~(a)
shows the result for cosine similarities of
the NMF basis vectors $\bm{D}_k$ for the two successive
months of 2019-09 and 2019-10. \figref{fig:comp_cf}~(b)
shows the result for $\bm{S}_k$ for the same data.
In either of these results, one can see that the NMF components
are quite similar except only a few permutation of indices.

\clearpage
\section{Discussions}\label{sec:discuss}

Let us briefly discuss about several aspects that would be worth
further investigation.

First, while we succeeded to extract the NMF components and found that
the components have non-zero values only at a relatively small number
of users, we still did not identify those users by exploiting the
fact. It is quite likely the case that the extracted users must play
important roles in each of the NMF components, either of key
destinations or key sources. We attempted to identify a tiny fraction
of such users by matching the list of such users with the identity
given in Appendix~\ref{sec:identiy}, but the identification was not
sufficient in order to interpret the meaning of corresponding NMF
components.  Instead, such intra-day activities as shown in
\figref{fig:hist_tx_utc_freq} of \secref{sec:data_users} could give us
the geographical locations of those key users, possibly uncovering the
crypto flow in each NMF component at a global scale. This issue
remains to be investigated.

Second, even if the temporal change of the network in terms of the NMF
components has such a stable structure as found in
\figref{fig:comp_cf}, we noticed that there exists interesting change
of a few components in the same figure. A keen reader may have noticed
that the components $k=3,4,5,6$ at time $t$ are changed into among
themselves at time $t+1$, while the cosine similarities are close to
1. This means that the probabilities $r_k$ for those components were
changed from one month to the next. Also one can notice that the
optimal number of NMF components showed a slow variation during the
period (recall \tabref{tab:optnum} of
Appendix~\ref{sec:optnum}). These facts might give us a hint for how
to treat the temporal change of network by paying attention to those
slowing varying aspects.  Additionally, while we focused only on
regular users appearing everyday during the period under study, it
would be necessary to include the process of entry and exit of big
players.

Third, technically, we regarded the method of NMF as a probabilistic
model that shares the same stochastic process as in the probabilistic
latent semantic analysis (PLSA). As a bonus, we were able to employ
the latent Dirichlet allocation (LDA) and its known methods to
estimate the number of topics in the context of topic model, or the
number of NMF components in out context. In principle, one could start
with the full-fledged Bayesian framework in the LDA and its extension
and variations. It would be worth pursuing in this direction, which
is also related to the second point above, because there are
several studies on how to treat temporally changing topics of
documents in a long time-span.

Fourth, our methods in this paper can be easily applied to different
cryptoassets including Ethereum and XRP. We are aware of the paper
\cite{akky2021pc} in this volume, which is in a similar line of
study. It would be interesting to apply our methods to the data of
XRP.

Finally, it would be an extremely interesting problem how the crypto
flow among big players is related to the \textit{prices} in the
exchange markets with fiat currencies and also with other
cryptoassets. It is quite likely that the bubble/crash and their
precursors might force the big players to react during such turmoils
in a different way from tranquil periods. For example, exchanges need
to reallocate cryptoassets in the necessity of making a reservoir or
doing a release of cryptoassets under the risk.

\section{Summary}\label{sec:summary}

Our purpose in this study on the cryptoasset of Bitcoin is to
understand the structure and temporal change of crypto flow among big
players. We compiled all the transactions contained in the blockchain
of Bitcoin cryptoasset from its genesis to the year 2020, identified
users from anonymous addresses, and constructed snapshots of networks
comprising of users as nodes and links as crypto flow among the
users. While the whole network is huge, we extracted sub-networks by
focusing on regular users who appeared persistently during a certain
period. Specifically, we extracted monthly snapshots during the year
of 2019, and selected roughly 500 regular users.

We first analyzed the bow-tie structure from the binary
relationship of flow, and then performed the Hodge decomposition
based on the strength of flow defined by frequencies and amounts,
in order to locate users in the upstream, downstream, and core
of the entire crypto flow. We found that the bow-tie structure
is stable during the period, implying that those regular users
have different roles in the crypto flow.

Then, to reveal important ingredients hidden in the flow,
we employ the method of non-negative matrix factorization (NMF)
to extract a set of principal components. We discussed that
the NMF method can be regarded as a probabilistic model,
which is equivalent to a probabilistic latent semantic analysis
and its typical model of latent Dirichlet allocation.
This observation brought us a method to estimate an optimal number
of NMF components, which turned out to be a dozen or so.
We found that the NMF components have non-zero values
corresponding to a limited number of users, telling us
their roles of destinations or sources of the crypto flow.
Additionally, we found that the NMF components are quite
stable in the temporal change for the time-scale of months.

There remain several points including the further investigation
on the users contained in those NMF components, a treatment
of temporally changing network, and technically interesting
issues to be pursued in the future.

\section*{Acknowledgment}

We would like to thank Hideaki Aoyama, Yuichi Ikeda, and Hiwon Yoon
for discussions, Hiroshi Iyetomi for solving a technical issue of
Hodge decomposition, Itsuki Noda for clarifying his application of NMF
on transportation data, Takeaki Uno for an efficient algorithm to
identify users, Shinya Kawata and Wajun Kawai for technical
assistance. This work is supported by JSPS KAKENHI Grant Numbers,
19K22032 and 20H02391, by the Nomura Foundation (Grants for Social
Science), and also by Kyoto University and Ripple's collaboration
scheme.


\appendix
\clearpage
\section{Identity of Users of Type A}\label{sec:identiy}

\texttt{WalletExplorer.com} \cite{walletexplorer} is a web site
providing information about identity of addresses in Bitcoin
blockchain. The site merges addresses together, if they are part of
the same wallet, and also identifies wallets with actual
names. According to the site, the method to merge addresses is:
\begin{quotation}\itshape\setlength\parindent{0pt}\noindent
  Just a basic algorithm is used to determine wallet
  addresses. Addresses are merged together, if they are co-spent in
  one transaction. So if addresses A and B are co-spent in transaction
  T1, and addresses B and C are co-spent in transaction T2, all
  addresses A, B and C will be part of one wallet.

  Sometimes, an address belongs to some service but it was never
  co-spent with others. Then that address stays unnamed. It is
  typically more often at addresses with higher amount (as there is no
  need to co-spending).
\end{quotation}
This method is precisely the same as \cite{reid2013aab}, which is the one
we employed in \secref{sec:data_users}.
In addition, the identification of actual names is done by
\texttt{WalletExplorer.com} as follows:
\begin{quotation}\itshape\setlength\parindent{0pt}\noindent
  In most of the cases, I registered to service, made transaction(s)
  and saw which wallet bitcoins were merged with, or from which wallet
  it was withdrawn.

  There is probably no easier way how to discover names other than
  this.

  Please note that the name database is not updated, so it does not
  contain newer exchanges (or newer wallets of existing exchanges).
\end{quotation}
We matched our data with the one in \cite{walletexplorer} to obtain
the identity and additional attributes of users of type A (see
\secref{sec:data_users} for the type). \tabref{tab:walletexplorer_cat}
is the classification into exchanges, services, gambling, historic,
and mining pools.  \tabref{tab:walletexplorer_country} shows the list
of countries that exchanges belong to.  \tabref{tab:walletexplorer} is
the complete list of this matching.

\begin{table}[h]
  \centering
  \caption{Classification of identified users (compiled from \cite{walletexplorer})}
  \label{tab:walletexplorer_cat}
  \begin{tabular}{lrl}
    \toprule
    Classification & \#Users & Examples \\
    \midrule
    Exchanges & 84 & \texttt{Bittrex.com}, \texttt{Huobi.com}, \texttt{Bit-x.com}, \texttt{HitBtc.com} \\
    Old/Historic & 83 & \texttt{AgoraMarket}, \texttt{EvolutionMarket}, \texttt{SilkRoadMarketplace} \\
    Services/Others & 45 & \texttt{Xapo.com}, \texttt{ePay.info}, \texttt{Cubits.com} \\
    Gambling & 41 & \texttt{999Dice.com}, \texttt{CoinGaming.io}, \texttt{SatoshiMines.com} \\
    Pools &  11 & \texttt{BTCCPool}, \texttt{SlushPool.com}, \texttt{BitMinter.com} \\
    \midrule
    Total & 264 & --- \\
    \bottomrule
  \end{tabular}
\end{table}

\captionof{table}{Countries of identified exchanges (compiled from \cite{walletexplorer})}
\label{tab:walletexplorer_country}
\begin{multicols}{3}
\begin{tabular}{lr}
\toprule
Country & \#Users \\
\midrule
China & 14 \\
UK & 13 \\
USA & 13 \\
Canada & 4 \\
Australia & 3 \\
Brazil & 3 \\
Singapore & 3 \\
Russia & 3 \\
Denmark & 2 \\
Finland & 2 \\
\end{tabular}

\columnbreak

\begin{tabular}{lr}
\midrule
Country & \#Users \\
\midrule
Mexico & 2 \\
Netherlands & 2 \\
Poland & 2 \\
Portugal & 2 \\
South Africa & 2 \\
Austria & 1 \\
Belize & 1 \\
Croatia & 1 \\
Czech & 1 \\
Germany & 1 \\
\end{tabular}

\columnbreak

\begin{tabular}{lr}
\midrule
Country & \#Users \\
\midrule
Iran & 1 \\
Korea & 1 \\
Luxembourg & 1 \\
Malta & 1 \\
Panama & 1 \\
Taiwan & 1 \\
Thailand & 1 \\
Vanuatu & 1 \\
Vietnam & 1 \\
\midrule
Total & 84 \\
\bottomrule
\end{tabular}

\end{multicols}

\begingroup
\scriptsize
\begin{longtable}{rcrrlll}
\caption{Identity of Users (compiled from \cite{walletexplorer})}
\label{tab:walletexplorer}
\\
\multicolumn{7}{l}{Definitions} \\
\multicolumn{7}{l}{\qquad No: sequential number} \\
\multicolumn{7}{l}{\qquad User ID: an arbitrarily but uniquely assigned IDs to each user in our data} \\
\multicolumn{7}{l}{\qquad \#Addr.~(1): number of addresses identified to each User ID in our data (rows are sorted by this column)} \\
\multicolumn{7}{l}{\qquad \#Addr.~(2): the same as (1) but provided by \texttt{WalletExplorer.com}
  (at the timing of writing)} \\
\multicolumn{7}{l}{} \\
\toprule
No. & User ID & \#Addr.~(1) & \#Addr.~(2) & Name & Category & Country \\
\midrule
\endfirsthead
\multicolumn{7}{l}{\textit{\tabref{tab:walletexplorer} continued from previous page}} \\
\midrule
No. & User ID & \#Addr.~(1) & \#Addr.~(2) & Name & Category & Country \\
\midrule
\endhead
\midrule
\multicolumn{7}{r}{\textit{Continue to next page}} \\
\endfoot
\bottomrule
\multicolumn{7}{r}{\textit{End of \tabref{tab:walletexplorer}}} \\
\endlastfoot

1 & \texttt{0000000000} & 18,913,420 & 1 & \texttt{Bit-x.com} & Exchanges & South Africa \\
2 & \texttt{0000000000} & 18,913,420 & 1 & \texttt{Xapo.com} & Services/Others & --- \\
3 & \texttt{0000000001} & 13,110,033 & 12,469,250 & \texttt{ePay.info} & Services/Others & --- \\
4 & \texttt{0000000002} & 5,302,867 & 1 & \texttt{Cryptopay.me} & Services/Others & --- \\
5 & \texttt{0000000002} & 5,302,867 & 1 & \texttt{Cubits.com} & Services/Others & --- \\
6 & \texttt{0000000002} & 5,302,867 & 1 & \texttt{Luno.com} & Exchanges & South Africa \\
7 & \texttt{0000000002} & 5,302,867 & 1 & \texttt{VirWoX.com} & Exchanges & Austria \\
8 & \texttt{0000000002} & 5,302,867 & 1 & \texttt{Xapo.com} & Services/Others & --- \\
9 & \texttt{0000000002} & 5,302,867 & 487,776 & \texttt{CoinPayments.net} & Services/Others & --- \\
10 & \texttt{0000000006} & 2,180,236 & 1,488,034 & \texttt{Xapo.com} & Services/Others & --- \\
11 & \texttt{0000000010} & 1,630,483 & 1 & \texttt{Huobi.com} & Exchanges & China \\
12 & \texttt{0000000010} & 1,630,483 & 957,652 & \texttt{Cubits.com} & Services/Others & --- \\
13 & \texttt{0000000011} & 1,538,476 & 1,377,461 & \texttt{Bittrex.com} & Exchanges & USA \\
14 & \texttt{0000000012} & 1,413,904 & 1 & \texttt{Bitstamp.net} & Exchanges & Luxembourg \\
15 & \texttt{0000000014} & 1,043,379 & 279,697 & \texttt{Huobi.com} & Exchanges & China \\
16 & \texttt{0000000017} & 992,726 & 1 & \texttt{VirWoX.com} & Exchanges & Austria \\
17 & \texttt{0000000017} & 992,726 & 1 & \texttt{Xapo.com} & Services/Others & --- \\
18 & \texttt{0000000018} & 988,300 & 940,605 & \texttt{Poloniex.com} & Exchanges & USA \\
19 & \texttt{0000000022} & 912,950 & 133,020 & \texttt{AnxPro.com} & Exchanges & China \\
20 & \texttt{0000000022} & 912,950 & 770,486 & \texttt{CoinTrader.net} & Exchanges & Canada \\
21 & \texttt{0000000025} & 845,559 & 1 & \texttt{Cubits.com} & Services/Others & --- \\
22 & \texttt{0000000028} & 811,809 & 2 & \texttt{Cubits.com} & Services/Others & --- \\
23 & \texttt{0000000030} & 778,990 & 651,547 & \texttt{999Dice.com} & Gambling & --- \\
24 & \texttt{0000000035} & 682,176 & 1 & \texttt{MoonBit.co.in} & Services/Others & --- \\
25 & \texttt{0000000037} & 659,820 & 656,699 & \texttt{CoinGaming.io} & Gambling & --- \\
26 & \texttt{0000000038} & 631,985 & 1 & \texttt{Luno.com} & Exchanges & South Africa \\
27 & \texttt{0000000041} & 581,525 & 1 & \texttt{Cubits.com} & Services/Others & --- \\
28 & \texttt{0000000042} & 546,515 & 291,443 & \texttt{Luno.com} & Exchanges & South Africa \\
29 & \texttt{0000000043} & 523,330 & 522,056 & \texttt{LocalBitcoins.com} & Exchanges & Finland \\
30 & \texttt{0000000045} & 498,001 & 498,001 & \texttt{AgoraMarket} & Old/Historic & --- \\
31 & \texttt{0000000049} & 478,476 & 343,039 & \texttt{Bitstamp.net} & Exchanges & Luxembourg \\
32 & \texttt{0000000054} & 420,632 & 420,632 & \texttt{EvolutionMarket} & Old/Historic & --- \\
33 & \texttt{0000000055} & 412,338 & 249,883 & \texttt{Cryptonator.com} & Services/Others & --- \\
34 & \texttt{0000000057} & 398,349 & 1 & \texttt{Xapo.com} & Services/Others & --- \\
35 & \texttt{0000000060} & 377,140 & 305,518 & \texttt{Cryptopay.me} & Services/Others & --- \\
36 & \texttt{0000000061} & 372,753 & 372,753 & \texttt{SilkRoadMarketplace} & Old/Historic & --- \\
37 & \texttt{0000000063} & 350,036 & 350,036 & \texttt{SilkRoad2Market} & Old/Historic & --- \\
38 & \texttt{0000000069} & 341,160 & 61,103 & \texttt{MercadoBitcoin.com.br} & Exchanges & Brazil \\
39 & \texttt{0000000071} & 325,365 & 1 & \texttt{Xapo.com} & Services/Others & --- \\
40 & \texttt{0000000075} & 307,489 & 307,451 & \texttt{BTC-e.com} & Exchanges & Russia \\
41 & \texttt{0000000077} & 294,238 & 1 & \texttt{Xapo.com} & Services/Others & --- \\
42 & \texttt{0000000079} & 278,973 & 2 & \texttt{Cubits.com} & Services/Others & --- \\
43 & \texttt{0000000082} & 266,695 & 254,601 & \texttt{SatoshiMines.com} & Gambling & --- \\
44 & \texttt{0000000083} & 263,074 & 262,940 & \texttt{YoBit.net} & Exchanges & Russia \\
45 & \texttt{0000000086} & 250,920 & 191,689 & \texttt{Bitcoin.de} & Exchanges & Germany \\
46 & \texttt{0000000090} & 241,250 & 223,781 & \texttt{BitcoinFog} & Services/Others & --- \\
47 & \texttt{0000000093} & 238,480 & 238,476 & \texttt{Cex.io} & Exchanges & UK \\
48 & \texttt{0000000101} & 206,542 & 156,203 & \texttt{CoinJar.com} & Services/Others & --- \\
49 & \texttt{0000000107} & 197,164 & 197,155 & \texttt{NitrogenSports.eu} & Gambling & --- \\
50 & \texttt{0000000114} & 189,776 & 189,776 & \texttt{AlphaBayMarket} & Services/Others & --- \\
51 & \texttt{0000000116} & 187,189 & 187,086 & \texttt{HitBtc.com} & Exchanges & UK \\
52 & \texttt{0000000120} & 186,000 & 186,000 & \texttt{BitPay.com} & Services/Others & --- \\
53 & \texttt{0000000135} & 168,885 & 1 & \texttt{Luno.com} & Exchanges & South Africa \\
54 & \texttt{0000000158} & 146,381 & 146,381 & \texttt{NucleusMarket} & Services/Others & --- \\
55 & \texttt{0000000159} & 145,978 & 1 & \texttt{Cryptonator.com} & Services/Others & --- \\
56 & \texttt{0000000159} & 145,978 & 1 & \texttt{Cubits.com} & Services/Others & --- \\
57 & \texttt{0000000169} & 140,594 & 1 & \texttt{Bitcoin.de} & Exchanges & Germany \\
58 & \texttt{0000000169} & 140,594 & 1 & \texttt{Cubits.com} & Services/Others & --- \\
59 & \texttt{0000000169} & 140,594 & 1 & \texttt{Poloniex.com} & Exchanges & USA \\
60 & \texttt{0000000176} & 134,559 & 134,559 & \texttt{Cryptsy.com} & Exchanges & USA \\
61 & \texttt{0000000182} & 131,979 & 1 & \texttt{Poloniex.com} & Exchanges & USA \\
62 & \texttt{0000000190} & 125,004 & 125,004 & \texttt{PocketDice.io} & Gambling & --- \\
63 & \texttt{0000000195} & 122,249 & 1 & \texttt{Bitstamp.net} & Exchanges & Luxembourg \\
64 & \texttt{0000000199} & 120,548 & 120,491 & \texttt{FortuneJack.com} & Gambling & --- \\
65 & \texttt{0000000201} & 119,119 & 119,065 & \texttt{AbraxasMarket} & Old/Historic & --- \\
66 & \texttt{0000000208} & 115,775 & 115,775 & \texttt{CoinKite.com} & Services/Others & --- \\
67 & \texttt{0000000210} & 114,458 & 114,458 & \texttt{Kraken.com} & Exchanges & USA \\
68 & \texttt{0000000216} & 109,798 & 1 & \texttt{Luno.com} & Exchanges & South Africa \\
69 & \texttt{0000000217} & 109,151 & 109,151 & \texttt{Instawallet.org} & Old/Historic & --- \\
70 & \texttt{0000000219} & 107,479 & 85,122 & \texttt{Bleutrade.com} & Exchanges & Brazil \\
71 & \texttt{0000000234} & 96,890 & 96,890 & \texttt{SecondsTrade.com} & Gambling & --- \\
72 & \texttt{0000000239} & 92,226 & 72,581 & \texttt{HolyTransaction.com} & Services/Others & --- \\
73 & \texttt{0000000244} & 90,473 & 44,416 & \texttt{CoinSpot.com.au} & Exchanges & Australia \\
74 & \texttt{0000000252} & 85,637 & 85,626 & \texttt{MintPal.com} & Old/Historic & --- \\
75 & \texttt{0000000253} & 85,566 & 84,679 & \texttt{Hashnest.com} & Exchanges & China \\
76 & \texttt{0000000259} & 83,723 & 83,517 & \texttt{BtcTrade.com} & Exchanges & China \\
77 & \texttt{0000000263} & 82,695 & 27,343 & \texttt{BTCJam.com} & Services/Others & --- \\
78 & \texttt{0000000267} & 80,987 & 73,965 & \texttt{OKCoin.com} & Exchanges & China \\
79 & \texttt{0000000269} & 80,086 & 1 & \texttt{Poloniex.com} & Exchanges & USA \\
80 & \texttt{0000000270} & 79,712 & 79,712 & \texttt{Bter.com} & Exchanges & China \\
81 & \texttt{0000000274} & 78,849 & 78,849 & \texttt{BitZino.com} & Gambling & --- \\
82 & \texttt{0000000278} & 78,119 & 78,048 & \texttt{OKCoin.com} & Exchanges & China \\
83 & \texttt{0000000282} & 76,997 & 1 & \texttt{Cubits.com} & Services/Others & --- \\
84 & \texttt{0000000282} & 76,997 & 52,925 & \texttt{BitoEX.com} & Services/Others & --- \\
85 & \texttt{0000000286} & 74,602 & 74,602 & \texttt{Rollin.io} & Gambling & --- \\
86 & \texttt{0000000300} & 68,956 & 67,795 & \texttt{CloudBet.com} & Gambling & --- \\
87 & \texttt{0000000302} & 68,658 & 55,098 & \texttt{VirWoX.com} & Exchanges & Austria \\
88 & \texttt{0000000307} & 67,114 & 1 & \texttt{Cubits.com} & Services/Others & --- \\
89 & \texttt{0000000312} & 66,748 & 1 & \texttt{Luno.com} & Exchanges & South Africa \\
90 & \texttt{0000000318} & 65,367 & 1 & \texttt{VirWoX.com} & Exchanges & Austria \\
91 & \texttt{0000000325} & 64,803 & 64,803 & \texttt{BTCC.com} & Exchanges & China \\
92 & \texttt{0000000347} & 57,770 & 57,753 & \texttt{MaiCoin.com} & Exchanges & Taiwan \\
93 & \texttt{0000000350} & 56,952 & 56,952 & \texttt{BTCCPool} & Pools & --- \\
94 & \texttt{0000000358} & 55,757 & 55,757 & \texttt{PandoraOpenMarket} & Old/Historic & --- \\
95 & \texttt{0000000359} & 55,703 & 1 & \texttt{MercadoBitcoin.com.br} & Exchanges & Brazil \\
96 & \texttt{0000000362} & 55,167 & 55,167 & \texttt{Paxful.com} & Exchanges & USA \\
97 & \texttt{0000000366} & 54,640 & 54,640 & \texttt{PrimeDice.com} & Gambling & --- \\
98 & \texttt{0000000370} & 53,639 & 53,639 & \texttt{SheepMarketplace} & Old/Historic & --- \\
99 & \texttt{0000000374} & 53,102 & 53,102 & \texttt{Cavirtex.com} & Exchanges & Canada \\
100 & \texttt{0000000387} & 50,878 & 50,878 & \texttt{BlackBankMarket} & Old/Historic & --- \\
101 & \texttt{0000000402} & 48,602 & 48,518 & \texttt{BX.in.th} & Exchanges & Thailand \\
102 & \texttt{0000000403} & 48,525 & 19,434 & \texttt{MoonBit.co.in} & Services/Others & --- \\
103 & \texttt{0000000407} & 48,178 & 33,372 & \texttt{HaoBTC.com} & Services/Others & --- \\
104 & \texttt{0000000412} & 47,295 & 47,281 & \texttt{Matbea.com} & Exchanges & UK \\
105 & \texttt{0000000444} & 42,750 & 41,150 & \texttt{SatoshiDice.com} & Gambling & --- \\
106 & \texttt{0000000447} & 42,327 & 41,866 & \texttt{BitcoinWallet.com} & Services/Others & --- \\
107 & \texttt{0000000468} & 40,270 & 1 & \texttt{Cubits.com} & Services/Others & --- \\
108 & \texttt{0000000480} & 39,013 & 1 & \texttt{VirWoX.com} & Exchanges & Austria \\
109 & \texttt{0000000482} & 38,880 & 1 & \texttt{Huobi.com} & Exchanges & China \\
110 & \texttt{0000000501} & 36,999 & 36,999 & \texttt{Justcoin.com} & Old/Historic & --- \\
111 & \texttt{0000000508} & 35,537 & 1 & \texttt{Bitcoin.de} & Exchanges & Germany \\
112 & \texttt{0000000508} & 35,537 & 28,567 & \texttt{SafeDice.com} & Gambling & --- \\
113 & \texttt{0000000511} & 35,453 & 35,453 & \texttt{McxNOW.com} & Old/Historic & --- \\
114 & \texttt{0000000512} & 35,433 & 35,433 & \texttt{C-Cex.com} & Exchanges & UK \\
115 & \texttt{0000000527} & 34,149 & 34,149 & \texttt{MiddleEarthMarketplace} & Old/Historic & --- \\
116 & \texttt{0000000533} & 33,436 & 33,389 & \texttt{Vircurex.com} & Exchanges & China \\
117 & \texttt{0000000537} & 32,823 & 32,823 & \texttt{Purse.io} & Services/Others & --- \\
118 & \texttt{0000000539} & 32,701 & 32,701 & \texttt{SatoshiBet.com} & Gambling & --- \\
119 & \texttt{0000000542} & 32,017 & 27,693 & \texttt{SwCPoker.eu} & Gambling & --- \\
120 & \texttt{0000000545} & 31,940 & 30,755 & \texttt{BitBargain.co.uk} & Exchanges & UK \\
121 & \texttt{0000000556} & 30,965 & 30,965 & \texttt{SealsWithClubs.eu} & Old/Historic & --- \\
122 & \texttt{0000000559} & 30,624 & 22,599 & \texttt{BlockTrades.us} & Exchanges & USA \\
123 & \texttt{0000000562} & 30,251 & 16,052 & \texttt{CoinMotion.com} & Exchanges & Finland \\
124 & \texttt{0000000563} & 30,187 & 30,183 & \texttt{OkLink.com} & Services/Others & --- \\
125 & \texttt{0000000571} & 29,256 & 29,256 & \texttt{Huobi.com} & Exchanges & China \\
126 & \texttt{0000000591} & 27,653 & 23,515 & \texttt{Bit-x.com} & Exchanges & South Africa \\
127 & \texttt{0000000605} & 26,622 & 26,622 & \texttt{BtcDice.com} & Old/Historic & --- \\
128 & \texttt{0000000614} & 26,013 & 24,204 & \texttt{BitBay.net} & Exchanges & Poland \\
129 & \texttt{0000000615} & 25,960 & 25,960 & \texttt{Betcoin.ag} & Gambling & --- \\
130 & \texttt{0000000626} & 25,257 & 1 & \texttt{Cubits.com} & Services/Others & --- \\
131 & \texttt{0000000649} & 23,841 & 20,866 & \texttt{Paymium.com} & Services/Others & --- \\
132 & \texttt{0000000672} & 22,597 & 22,597 & \texttt{Loanbase.com} & Services/Others & --- \\
133 & \texttt{0000000676} & 22,304 & 22,304 & \texttt{Coinroll.com} & Gambling & --- \\
134 & \texttt{0000000687} & 21,693 & 21,693 & \texttt{FaucetBOX.com} & Services/Others & --- \\
135 & \texttt{0000000690} & 21,640 & 10,439 & \texttt{CoinHako.com} & Exchanges & Singapore \\
136 & \texttt{0000000717} & 20,484 & 18,001 & \texttt{FYBSG.com} & Exchanges & Singapore \\
137 & \texttt{0000000747} & 19,810 & 19,605 & \texttt{TheRockTrading.com} & Exchanges & Malta \\
138 & \texttt{0000000762} & 18,997 & 18,997 & \texttt{BlueSkyMarketplace} & Old/Historic & --- \\
139 & \texttt{0000000774} & 18,489 & 18,489 & \texttt{Crypto-Games.net} & Gambling & --- \\
140 & \texttt{0000000794} & 17,705 & 17,705 & \texttt{Coin-Swap.net} & Old/Historic & --- \\
141 & \texttt{0000000804} & 17,531 & 1 & \texttt{Luno.com} & Exchanges & South Africa \\
142 & \texttt{0000000869} & 15,965 & 15,965 & \texttt{AnoniBet.com} & Gambling & --- \\
143 & \texttt{0000000874} & 15,905 & 15,905 & \texttt{ChangeTip.com} & Services/Others & --- \\
144 & \texttt{0000000881} & 15,757 & 15,757 & \texttt{Bitmit.net} & Old/Historic & --- \\
145 & \texttt{0000000891} & 15,495 & 15,495 & \texttt{CoinApult.com} & Services/Others & --- \\
146 & \texttt{0000000902} & 15,260 & 15,260 & \texttt{BtcMarkets.net} & Exchanges & Australia \\
147 & \texttt{0000000926} & 14,566 & 14,566 & \texttt{Inputs.io} & Old/Historic & --- \\
148 & \texttt{0000000934} & 14,394 & 1 & \texttt{Huobi.com} & Exchanges & China \\
149 & \texttt{0000000959} & 13,713 & 11,210 & \texttt{Vaultoro.com} & Exchanges & UK \\
150 & \texttt{0000001006} & 12,486 & 12,486 & \texttt{CryptoStocks.com} & Services/Others & --- \\
151 & \texttt{0000001007} & 12,456 & 12,456 & \texttt{BitAces.me} & Old/Historic & --- \\
152 & \texttt{0000001071} & 11,221 & 11,221 & \texttt{Coins-e.com} & Exchanges & Canada \\
153 & \texttt{0000001072} & 11,220 & 11,220 & \texttt{Igot.com} & Exchanges & Belize \\
154 & \texttt{0000001093} & 10,901 & 10,900 & \texttt{SatoshiRoulette.com} & Gambling & --- \\
155 & \texttt{0000001177} & 9,667 & 1 & \texttt{Bittrex.com} & Exchanges & USA \\
156 & \texttt{0000001199} & 9,512 & 9,512 & \texttt{Crypto-Trade.com} & Old/Historic & --- \\
157 & \texttt{0000001235} & 9,165 & 9,165 & \texttt{Cryptorush.in} & Old/Historic & --- \\
158 & \texttt{0000001240} & 9,122 & 9,122 & \texttt{BTCOracle.com} & Gambling & --- \\
159 & \texttt{0000001255} & 8,967 & 8,967 & \texttt{Genesis-Mining.com} & Services/Others & --- \\
160 & \texttt{0000001313} & 8,430 & 8,430 & \texttt{Exmo.com} & Exchanges & UK \\
161 & \texttt{0000001325} & 8,371 & 4,343 & \texttt{SlushPool.com} & Pools & --- \\
162 & \texttt{0000001355} & 8,120 & 8,120 & \texttt{VaultOfSatoshi.com} & Old/Historic & --- \\
163 & \texttt{0000001368} & 8,032 & 8,032 & \texttt{BitcoinVideoCasino.com} & Gambling & --- \\
164 & \texttt{0000001393} & 7,865 & 7,865 & \texttt{BTCGuild.com} & Old/Historic & --- \\
165 & \texttt{0000001395} & 7,857 & 7,766 & \texttt{Peerbet.org} & Gambling & --- \\
166 & \texttt{0000001397} & 7,848 & 7,848 & \texttt{796.com} & Exchanges & Vanuatu \\
167 & \texttt{0000001421} & 7,585 & 7,585 & \texttt{Btc38.com} & Exchanges & UK \\
168 & \texttt{0000001438} & 7,479 & 7,479 & \texttt{Betcoins.net} & Old/Historic & --- \\
169 & \texttt{0000001497} & 7,109 & 7,109 & \texttt{LiteBit.eu} & Exchanges & Netherlands \\
170 & \texttt{0000001568} & 6,788 & 6,207 & \texttt{Bitbond.com} & Services/Others & --- \\
171 & \texttt{0000001588} & 6,669 & 5,369 & \texttt{HappyCoins.com} & Exchanges & Netherlands \\
172 & \texttt{0000001631} & 6,477 & 6,477 & \texttt{Bitcoin-Roulette.com} & Old/Historic & --- \\
173 & \texttt{0000001666} & 6,309 & 6,309 & \texttt{AllCoin.com} & Old/Historic & --- \\
174 & \texttt{0000001685} & 6,242 & 6,242 & \texttt{Coin.mx} & Old/Historic & --- \\
175 & \texttt{0000001739} & 6,009 & 4,726 & \texttt{LakeBTC.com} & Exchanges & China \\
176 & \texttt{0000001750} & 5,966 & 5,966 & \texttt{777Coin.com} & Gambling & --- \\
177 & \texttt{0000001760} & 5,934 & 5,934 & \texttt{GHash.io} & Pools & --- \\
178 & \texttt{0000001796} & 5,762 & 5,762 & \texttt{DoctorDMarket} & Services/Others & --- \\
179 & \texttt{0000001862} & 5,481 & 5,481 & \texttt{Coinomat.com} & Exchanges & UK \\
180 & \texttt{0000001916} & 5,297 & 5,295 & \texttt{Coinmate.io} & Exchanges & UK \\
181 & \texttt{0000002012} & 5,024 & 5,024 & \texttt{BitVC.com} & Exchanges & China \\
182 & \texttt{0000002024} & 4,996 & 4,875 & \texttt{SatoshiCircle.com} & Gambling & --- \\
183 & \texttt{0000002039} & 4,953 & 1 & \texttt{Luno.com} & Exchanges & South Africa \\
184 & \texttt{0000002055} & 4,896 & 4,896 & \texttt{MyBitcoin.com} & Old/Historic & --- \\
185 & \texttt{0000002098} & 4,775 & 4,775 & \texttt{AllCrypt.com} & Old/Historic & --- \\
186 & \texttt{0000002155} & 4,629 & 4,629 & \texttt{GermanPlazaMarket} & Services/Others & --- \\
187 & \texttt{0000002169} & 4,605 & 4,605 & \texttt{MasterXchange.com} & Old/Historic & --- \\
188 & \texttt{0000002183} & 4,551 & 4,272 & \texttt{CoinCafe.com} & Exchanges & USA \\
189 & \texttt{0000002192} & 4,530 & 4,242 & \texttt{BitKonan.com} & Exchanges & Croatia \\
190 & \texttt{0000002197} & 4,516 & 4,516 & \texttt{QuadrigaCX.com} & Exchanges & Canada \\
191 & \texttt{0000002216} & 4,451 & 4,451 & \texttt{BitElfin.com} & Old/Historic & --- \\
192 & \texttt{0000002243} & 4,377 & 4,377 & \texttt{OrderBook.net} & Exchanges & USA \\
193 & \texttt{0000002251} & 4,363 & 3,800 & \texttt{SpectroCoin.com} & Exchanges & UK \\
194 & \texttt{0000002259} & 4,354 & 4,354 & \texttt{Bitcurex.com} & Exchanges & Poland \\
195 & \texttt{0000002265} & 4,338 & 4,338 & \texttt{Coinichiwa.com} & Gambling & --- \\
196 & \texttt{0000002281} & 4,292 & 4,277 & \texttt{Betcoin.tm} & Gambling & --- \\
197 & \texttt{0000002305} & 4,232 & 4,227 & \texttt{MeXBT.com} & Exchanges & Mexico \\
198 & \texttt{0000002403} & 3,999 & 3,999 & \texttt{Bitfinex.com} & Exchanges & China \\
199 & \texttt{0000002424} & 3,947 & 1,571 & \texttt{CoinVault} & Old/Historic & --- \\
200 & \texttt{0000002464} & 3,861 & 3,861 & \texttt{BetsOfBitco.in} & Old/Historic & --- \\
201 & \texttt{0000002471} & 3,840 & 3,840 & \texttt{JetWin.com} & Gambling & --- \\
202 & \texttt{0000002587} & 3,607 & 3,607 & \texttt{BitZillions.com} & Gambling & --- \\
203 & \texttt{0000002617} & 3,545 & 3,543 & \texttt{Korbit.co.kr} & Exchanges & Korea \\
204 & \texttt{0000002661} & 3,485 & 3,485 & \texttt{BTCPop.co} & Services/Others & --- \\
205 & \texttt{0000002849} & 3,216 & 2,181 & \texttt{YABTCL.com} & Gambling & --- \\
206 & \texttt{0000002922} & 3,121 & 3,121 & \texttt{BIToomBa.com} & Old/Historic & --- \\
207 & \texttt{0000002952} & 3,086 & 3,086 & \texttt{BitYes.com} & Old/Historic & --- \\
208 & \texttt{0000002965} & 3,075 & 2,640 & \texttt{BetMoose.com} & Gambling & --- \\
209 & \texttt{0000003031} & 2,979 & 2,978 & \texttt{CoinURL.com} & Services/Others & --- \\
210 & \texttt{0000003139} & 2,829 & 2,829 & \texttt{CannabisRoadMarket} & Old/Historic & --- \\
211 & \texttt{0000003195} & 2,760 & 2,760 & \texttt{Ice-Dice.com} & Old/Historic & --- \\
212 & \texttt{0000003211} & 2,744 & 2,744 & \texttt{ChBtc.com} & Exchanges & China \\
213 & \texttt{0000003249} & 2,713 & 2,713 & \texttt{CoinArch.com} & Exchanges & Singapore \\
214 & \texttt{0000003310} & 2,645 & 2,645 & \texttt{Comkort.com} & Old/Historic & --- \\
215 & \texttt{0000003340} & 2,618 & 2,618 & \texttt{BitNZ.com} & Services/Others & --- \\
216 & \texttt{0000003348} & 2,614 & 2,614 & \texttt{CleverCoin.com} & Exchanges & USA \\
217 & \texttt{0000003388} & 2,575 & 2,575 & \texttt{CoinMkt.com} & Old/Historic & --- \\
218 & \texttt{0000003637} & 2,355 & 2,355 & \texttt{DiceBitco.in} & Old/Historic & --- \\
219 & \texttt{0000003756} & 2,276 & 2,276 & \texttt{BitcoinVietnam.com.vn} & Exchanges & Vietnam \\
220 & \texttt{0000003840} & 2,221 & 2,221 & \texttt{Indacoin.com} & Exchanges & UK \\
221 & \texttt{0000004134} & 2,023 & 2,023 & \texttt{BitClix.com} & Services/Others & --- \\
222 & \texttt{0000004187} & 1,992 & 1,992 & \texttt{Coin-Sweeper.com} & Old/Historic & --- \\
223 & \texttt{0000004271} & 1,948 & 1,948 & \texttt{GoCelery.com} & Services/Others & --- \\
224 & \texttt{0000004570} & 1,812 & 1,812 & \texttt{Playt.in} & Old/Historic & --- \\
225 & \texttt{0000004580} & 1,804 & 1,796 & \texttt{Bitcash.cz} & Old/Historic & --- \\
226 & \texttt{0000004586} & 1,802 & 1,802 & \texttt{CampBX.com} & Exchanges & USA \\
227 & \texttt{0000004817} & 1,713 & 1,713 & \texttt{BTCLend.org} & Services/Others & --- \\
228 & \texttt{0000004840} & 1,704 & 1,704 & \texttt{CoinChimp.com} & Exchanges & Russia \\
229 & \texttt{0000004863} & 1,699 & 1,699 & \texttt{BtcExchange.ro} & Old/Historic & --- \\
230 & \texttt{0000004882} & 1,690 & 1,690 & \texttt{AdmiralCoin.com} & Old/Historic & --- \\
231 & \texttt{0000005002} & 1,643 & 1,643 & \texttt{Bitcoinica.com} & Old/Historic & --- \\
232 & \texttt{0000005121} & 1,594 & 1,594 & \texttt{Gatecoin.com} & Exchanges & China \\
233 & \texttt{0000005399} & 1,508 & 1,508 & \texttt{BetChain.com-old} & Gambling & --- \\
234 & \texttt{0000005547} & 1,471 & 1,471 & \texttt{BabylonMarket} & Old/Historic & --- \\
235 & \texttt{0000005637} & 1,443 & 1,443 & \texttt{HelixMixer} & Services/Others & --- \\
236 & \texttt{0000006104} & 1,314 & 1,314 & \texttt{Bylls.com} & Services/Others & --- \\
237 & \texttt{0000006381} & 1,246 & 1,246 & \texttt{Btcst.com-pirateat40} & Old/Historic & --- \\
238 & \texttt{0000006688} & 1,189 & 1,189 & \texttt{PocketRocketsCasino.eu} & Old/Historic & --- \\
239 & \texttt{0000006697} & 1,188 & 1,188 & \texttt{Bitso.com} & Exchanges & Mexico \\
240 & \texttt{0000006747} & 1,178 & 1,178 & \texttt{BTCt.com} & Old/Historic & --- \\
241 & \texttt{0000006753} & 1,176 & 1,176 & \texttt{DaDice.com} & Old/Historic & --- \\
242 & \texttt{0000007201} & 1,095 & 1,095 & \texttt{Cryptonit.net} & Exchanges & UK \\
243 & \texttt{0000007312} & 1,076 & 1,076 & \texttt{BitStarz.com} & Gambling & --- \\
244 & \texttt{0000007547} & 1,040 & 1,040 & \texttt{Ccedk.com} & Exchanges & Denmark \\
245 & \texttt{0000007673} & 1,020 & 1,020 & \texttt{Satoshi-Karoshi.com} & Gambling & --- \\
246 & \texttt{0000007783} & 1,005 & 1 & \texttt{VirWoX.com} & Exchanges & Austria \\
247 & \texttt{0000008133} & 978 & 978 & \texttt{Just-Dice.com} & Old/Historic & --- \\
248 & \texttt{0000008213} & 968 & 968 & \texttt{CryptoLocker} & Old/Historic & --- \\
249 & \texttt{0000008232} & 965 & 965 & \texttt{GreenRoadMarket} & Services/Others & --- \\
250 & \texttt{0000008325} & 953 & 953 & \texttt{CoinRoyale.com} & Gambling & --- \\
251 & \texttt{0000008352} & 950 & 950 & \texttt{CryptoBounty.com} & Old/Historic & --- \\
252 & \texttt{0000009025} & 876 & 876 & \texttt{1Coin.com} & Exchanges & China \\
253 & \texttt{0000009100} & 869 & 679 & \texttt{Coingi.com} & Exchanges & Panama \\
254 & \texttt{0000010744} & 746 & 746 & \texttt{BitcoinWeBank.com} & Old/Historic & --- \\
255 & \texttt{0000010833} & 741 & 741 & \texttt{EmpoEX.com} & Exchanges & USA \\
256 & \texttt{0000011035} & 729 & 729 & \texttt{FairProof.com} & Gambling & --- \\
257 & \texttt{0000011546} & 700 & 700 & \texttt{UseCryptos.com} & Exchanges & Portugal \\
258 & \texttt{0000011749} & 686 & 1 & \texttt{Cubits.com} & Services/Others & --- \\
259 & \texttt{0000011857} & 681 & 1 & \texttt{VirWoX.com} & Exchanges & Austria \\
260 & \texttt{0000012513} & 670 & 669 & \texttt{AntPool.com} & Pools & --- \\
261 & \texttt{0000012569} & 668 & 668 & \texttt{Coinbroker.io} & Exchanges & USA \\
262 & \texttt{0000012639} & 665 & 665 & \texttt{UpDown.BT} & Old/Historic & --- \\
263 & \texttt{0000013022} & 648 & 648 & \texttt{DiceNow.com} & Gambling & --- \\
264 & \texttt{0000013139} & 643 & 643 & \texttt{Dagensia.eu} & Old/Historic & --- \\
265 & \texttt{0000013870} & 614 & 614 & \texttt{WatchMyBit.com} & Services/Others & --- \\
266 & \texttt{0000014175} & 604 & 604 & \texttt{MPEx.co} & Old/Historic & --- \\
267 & \texttt{0000014703} & 593 & 593 & \texttt{Banx.io} & Exchanges & USA \\
268 & \texttt{0000015353} & 572 & 572 & \texttt{CloudHashing.com} & Old/Historic & --- \\
269 & \texttt{0000015549} & 565 & 565 & \texttt{Eligius.st} & Pools & --- \\
270 & \texttt{0000016775} & 527 & 527 & \texttt{Europex.eu} & Old/Historic & --- \\
271 & \texttt{0000017189} & 515 & 515 & \texttt{EveryDice.com} & Old/Historic & --- \\
272 & \texttt{0000018179} & 499 & 499 & \texttt{Brawker.com} & Old/Historic & --- \\
273 & \texttt{0000019489} & 473 & 471 & \texttt{10xBitco.in} & Old/Historic & --- \\
274 & \texttt{0000021056} & 440 & 1 & \texttt{Cubits.com} & Services/Others & --- \\
275 & \texttt{0000021084} & 440 & 439 & \texttt{BitMinter.com} & Pools & --- \\
276 & \texttt{0000021286} & 436 & 436 & \texttt{ExchangeMyCoins.com} & Exchanges & Denmark \\
277 & \texttt{0000023185} & 402 & 402 & \texttt{BW.com} & Pools & --- \\
278 & \texttt{0000023304} & 400 & 400 & \texttt{Chainroll.com} & Old/Historic & --- \\
279 & \texttt{0000024708} & 390 & 390 & \texttt{DiceCoin.io} & Gambling & --- \\
280 & \texttt{0000025249} & 382 & 382 & \texttt{FoxBit.com.br} & Exchanges & Brazil \\
281 & \texttt{0000025389} & 380 & 380 & \texttt{PonziCoin.co} & Old/Historic & --- \\
282 & \texttt{0000025554} & 377 & 377 & \texttt{Birwo.com-old} & Old/Historic & --- \\
283 & \texttt{0000027509} & 351 & 338 & \texttt{Zyado.com} & Exchanges & Portugal \\
284 & \texttt{0000028306} & 341 & 341 & \texttt{SuzukiDice.com} & Old/Historic & --- \\
285 & \texttt{0000028897} & 334 & 1 & \texttt{Cubits.com} & Services/Others & --- \\
286 & \texttt{0000032828} & 297 & 297 & \texttt{KnCMiner.com} & Pools & --- \\
287 & \texttt{0000033285} & 293 & 293 & \texttt{BitcoinPokerTables.com} & Gambling & --- \\
288 & \texttt{0000034753} & 280 & 280 & \texttt{Polmine.pl} & Old/Historic & --- \\
289 & \texttt{0000034912} & 279 & 279 & \texttt{MineField.BitcoinLab.org} & Gambling & --- \\
290 & \texttt{0000045961} & 234 & 234 & \texttt{SmenarnaBitcoin.cz} & Old/Historic & --- \\
291 & \texttt{0000062902} & 195 & 195 & \texttt{Dgex.com} & Old/Historic & --- \\
292 & \texttt{0000064830} & 190 & 190 & \texttt{BitLaunder.com} & Services/Others & --- \\
293 & \texttt{0000072390} & 170 & 170 & \texttt{BitMillions.com} & Old/Historic & --- \\
294 & \texttt{0000074148} & 166 & 1 & \texttt{Cubits.com} & Services/Others & --- \\
295 & \texttt{0000078381} & 157 & 1 & \texttt{Cubits.com} & Services/Others & --- \\
296 & \texttt{0000100904} & 125 & 125 & \texttt{Vic-Socks.to} & Services/Others & --- \\
297 & \texttt{0000108438} & 117 & 117 & \texttt{Gatecoin.com} & Exchanges & China \\
298 & \texttt{0000109665} & 116 & 1 & \texttt{VirWoX.com} & Exchanges & Austria \\
299 & \texttt{0000114095} & 112 & 112 & \texttt{Bitcoin-24.com} & Old/Historic & --- \\
300 & \texttt{0000157019} & 96 & 1 & \texttt{Cubits.com} & Services/Others & --- \\
301 & \texttt{0000180141} & 85 & 1 & \texttt{Poloniex.com} & Exchanges & USA \\
302 & \texttt{0000200218} & 77 & 77 & \texttt{BetcoinDice.tm} & Old/Historic & --- \\
303 & \texttt{0000206094} & 75 & 75 & \texttt{Bitfury.org} & Pools & --- \\
304 & \texttt{0000265341} & 59 & 1 & \texttt{Cubits.com} & Services/Others & --- \\
305 & \texttt{0000296609} & 53 & 53 & \texttt{50BTC.com} & Old/Historic & --- \\
306 & \texttt{0000302266} & 52 & 1 & \texttt{Cubits.com} & Services/Others & --- \\
307 & \texttt{0000302626} & 52 & 1 & \texttt{Cubits.com} & Services/Others & --- \\
308 & \texttt{0000332619} & 49 & 49 & \texttt{BtcEur.eu} & Old/Historic & --- \\
309 & \texttt{0000417570} & 40 & 1 & \texttt{VirWoX.com} & Exchanges & Austria \\
310 & \texttt{0000434094} & 38 & 1 & \texttt{Luno.com} & Exchanges & South Africa \\
311 & \texttt{0000482184} & 35 & 35 & \texttt{SecureVPN.to} & Services/Others & --- \\
312 & \texttt{0000489034} & 34 & 1 & \texttt{Xapo.com} & Services/Others & --- \\
313 & \texttt{0000539744} & 32 & 32 & \texttt{MinersCenter.com} & Old/Historic & --- \\
314 & \texttt{0000661573} & 28 & 28 & \texttt{DiceOnCrack.com} & Old/Historic & --- \\
315 & \texttt{0000727720} & 26 & 1 & \texttt{Cubits.com} & Services/Others & --- \\
316 & \texttt{0000881383} & 23 & 1 & \texttt{Bittrex.com} & Exchanges & USA \\
317 & \texttt{0001047894} & 20 & 20 & \texttt{ActionCrypto.com} & Old/Historic & --- \\
318 & \texttt{0001068642} & 20 & 1 & \texttt{Cubits.com} & Services/Others & --- \\
319 & \texttt{0001453639} & 15 & 1 & \texttt{CoinMotion.com} & Exchanges & Finland \\
320 & \texttt{0001482058} & 15 & 15 & \texttt{ASICMiner} & Old/Historic & --- \\
321 & \texttt{0001615111} & 14 & 14 & \texttt{Telco214} & Pools & --- \\
322 & \texttt{0001849904} & 12 & 12 & \texttt{BTradeAustralia.com} & Exchanges & Australia \\
323 & \texttt{0001949672} & 12 & 12 & \texttt{PinballCoin.com} & Old/Historic & --- \\
324 & \texttt{0002236961} & 11 & 1 & \texttt{MoonBit.co.in} & Services/Others & --- \\
325 & \texttt{0002437346} & 10 & 1 & \texttt{Cubits.com} & Services/Others & --- \\
326 & \texttt{0002652760} & 9 & 1 & \texttt{Xapo.com} & Services/Others & --- \\
327 & \texttt{0002867200} & 9 & 1 & \texttt{Cubits.com} & Services/Others & --- \\
328 & \texttt{0003042379} & 8 & 1 & \texttt{Xapo.com} & Services/Others & --- \\
329 & \texttt{0003656659} & 7 & 1 & \texttt{Cubits.com} & Services/Others & --- \\
330 & \texttt{0003733569} & 7 & 1 & \texttt{Luno.com} & Exchanges & South Africa \\
331 & \texttt{0004053833} & 7 & 7 & \texttt{SimpleCoin.cz} & Exchanges & Czech \\
332 & \texttt{0004615543} & 6 & 6 & \texttt{LuckyB.it} & Gambling & --- \\
333 & \texttt{0004719731} & 6 & 1 & \texttt{Cubits.com} & Services/Others & --- \\
334 & \texttt{0005083698} & 6 & 1 & \texttt{Cubits.com} & Services/Others & --- \\
335 & \texttt{0005134566} & 6 & 1 & \texttt{Cubits.com} & Services/Others & --- \\
336 & \texttt{0005316858} & 6 & 1 & \texttt{LakeBTC.com} & Exchanges & China \\
337 & \texttt{0006291300} & 5 & 1 & \texttt{Cryptopay.me} & Services/Others & --- \\
338 & \texttt{0006553479} & 5 & 1 & \texttt{Xapo.com} & Services/Others & --- \\
339 & \texttt{0006787887} & 5 & 1 & \texttt{Cubits.com} & Services/Others & --- \\
340 & \texttt{0006884235} & 5 & 5 & \texttt{Exchanging.ir} & Exchanges & Iran \\
341 & \texttt{0007872688} & 4 & 1 & \texttt{Cubits.com} & Services/Others & --- \\
342 & \texttt{0008064964} & 4 & 4 & \texttt{FoxBit.com.br} & Exchanges & Brazil \\
343 & \texttt{0008251183} & 4 & 1 & \texttt{Cubits.com} & Services/Others & --- \\
344 & \texttt{0009938959} & 4 & 1 & \texttt{Cubits.com} & Services/Others & --- \\
345 & \texttt{0010541278} & 4 & 1 & \texttt{Bitcoin.de} & Exchanges & Germany \\
346 & \texttt{0010933272} & 4 & 4 & \texttt{ePay.info} & Services/Others & --- \\
347 & \texttt{0011084834} & 4 & 1 & \texttt{Cubits.com} & Services/Others & --- \\
348 & \texttt{0011993584} & 3 & 1 & \texttt{Cubits.com} & Services/Others & --- \\
349 & \texttt{0013456456} & 3 & 3 & \texttt{EclipseMC.com} & Pools & --- \\
350 & \texttt{0014115232} & 3 & 1 & \texttt{Xapo.com} & Services/Others & --- \\
351 & \texttt{0014366385} & 3 & 1 & \texttt{Cubits.com} & Services/Others & --- \\
352 & \texttt{0014572029} & 3 & 1 & \texttt{StrongCoin.com-fee} & Services/Others & --- \\
353 & \texttt{0015020521} & 3 & 1 & \texttt{Poloniex.com} & Exchanges & USA \\
354 & \texttt{0015254424} & 3 & 1 & \texttt{Cubits.com} & Services/Others & --- \\
355 & \texttt{0017326362} & 3 & 1 & \texttt{Cubits.com} & Services/Others & --- \\
356 & \texttt{0018289896} & 3 & 1 & \texttt{Huobi.com} & Exchanges & China \\
357 & \texttt{0018738066} & 3 & 1 & \texttt{Cubits.com} & Services/Others & --- \\
358 & \texttt{0020356858} & 3 & 1 & \texttt{Poloniex.com} & Exchanges & USA \\
359 & \texttt{0021062591} & 3 & 1 & \texttt{Cubits.com} & Services/Others & --- \\
360 & \texttt{0021163228} & 3 & 3 & \texttt{ePay.info} & Services/Others & --- \\
361 & \texttt{0021447975} & 2 & 1 & \texttt{Cubits.com} & Services/Others & --- \\
362 & \texttt{0022923752} & 2 & 1 & \texttt{Cubits.com} & Services/Others & --- \\
363 & \texttt{0023037394} & 2 & 1 & \texttt{Cubits.com} & Services/Others & --- \\
364 & \texttt{0024428379} & 2 & 1 & \texttt{Cubits.com} & Services/Others & --- \\
365 & \texttt{0026028525} & 2 & 1 & \texttt{Cubits.com} & Services/Others & --- \\
366 & \texttt{0033718608} & 2 & 1 & \texttt{Xapo.com} & Services/Others & --- \\
367 & \texttt{0034424835} & 2 & 1 & \texttt{VirWoX.com} & Exchanges & Austria \\
368 & \texttt{0034774492} & 2 & 1 & \texttt{Cubits.com} & Services/Others & --- \\
369 & \texttt{0035023024} & 2 & 1 & \texttt{MercadoBitcoin.com.br} & Exchanges & Brazil \\
370 & \texttt{0035093871} & 2 & 1 & \texttt{Huobi.com} & Exchanges & China \\
371 & \texttt{0036244970} & 2 & 1 & \texttt{Cubits.com} & Services/Others & --- \\
372 & \texttt{0036353675} & 2 & 1 & \texttt{Cubits.com} & Services/Others & --- \\
373 & \texttt{0036389109} & 2 & 2 & \texttt{Dispenser.tf} & Old/Historic & --- \\
374 & \texttt{0039716556} & 2 & 1 & \texttt{Poloniex.com} & Exchanges & USA \\
375 & \texttt{0040013060} & 2 & 1 & \texttt{Cubits.com} & Services/Others & --- \\
376 & \texttt{0045999467} & 2 & 2 & \texttt{DeepBit.net} & Old/Historic & --- \\
377 & \texttt{0046840841} & 2 & 1 & \texttt{Cubits.com} & Services/Others & --- \\
378 & \texttt{0047562289} & 2 & 1 & \texttt{Cubits.com} & Services/Others & --- \\
379 & \texttt{0049075064} & 2 & 1 & \texttt{Cubits.com} & Services/Others & --- \\
380 & \texttt{0050709449} & 2 & 1 & \texttt{Poloniex.com} & Exchanges & USA \\
381 & \texttt{0051958689} & 2 & 2 & \texttt{CoinWorker.com} & Services/Others & --- \\
382 & \texttt{0052140982} & 2 & 1 & \texttt{Cubits.com} & Services/Others & --- \\
383 & \texttt{0053961615} & 2 & 1 & \texttt{Cubits.com} & Services/Others & --- \\
384 & \texttt{0055586746} & 2 & 1 & \texttt{Cubits.com} & Services/Others & --- \\
385 & \texttt{0056053704} & 2 & 1 & \texttt{Luno.com} & Exchanges & South Africa \\
386 & \texttt{0058665155} & 2 & 1 & \texttt{Xapo.com} & Services/Others & --- \\
387 & \texttt{0058696998} & 2 & 1 & \texttt{Cubits.com} & Services/Others & --- \\
388 & \texttt{0058767317} & 2 & 1 & \texttt{Cubits.com} & Services/Others & --- \\
389 & \texttt{0059011161} & 2 & 1 & \texttt{Bittrex.com} & Exchanges & USA \\

\end{longtable}
\endgroup

\clearpage
\section{Illustrating Adjacent Matrices of $G_t$ or $X_{sd}$}\label{sec:adj}

\begin{figure}[h]
  \centering
  \includegraphics[width=0.98\linewidth]{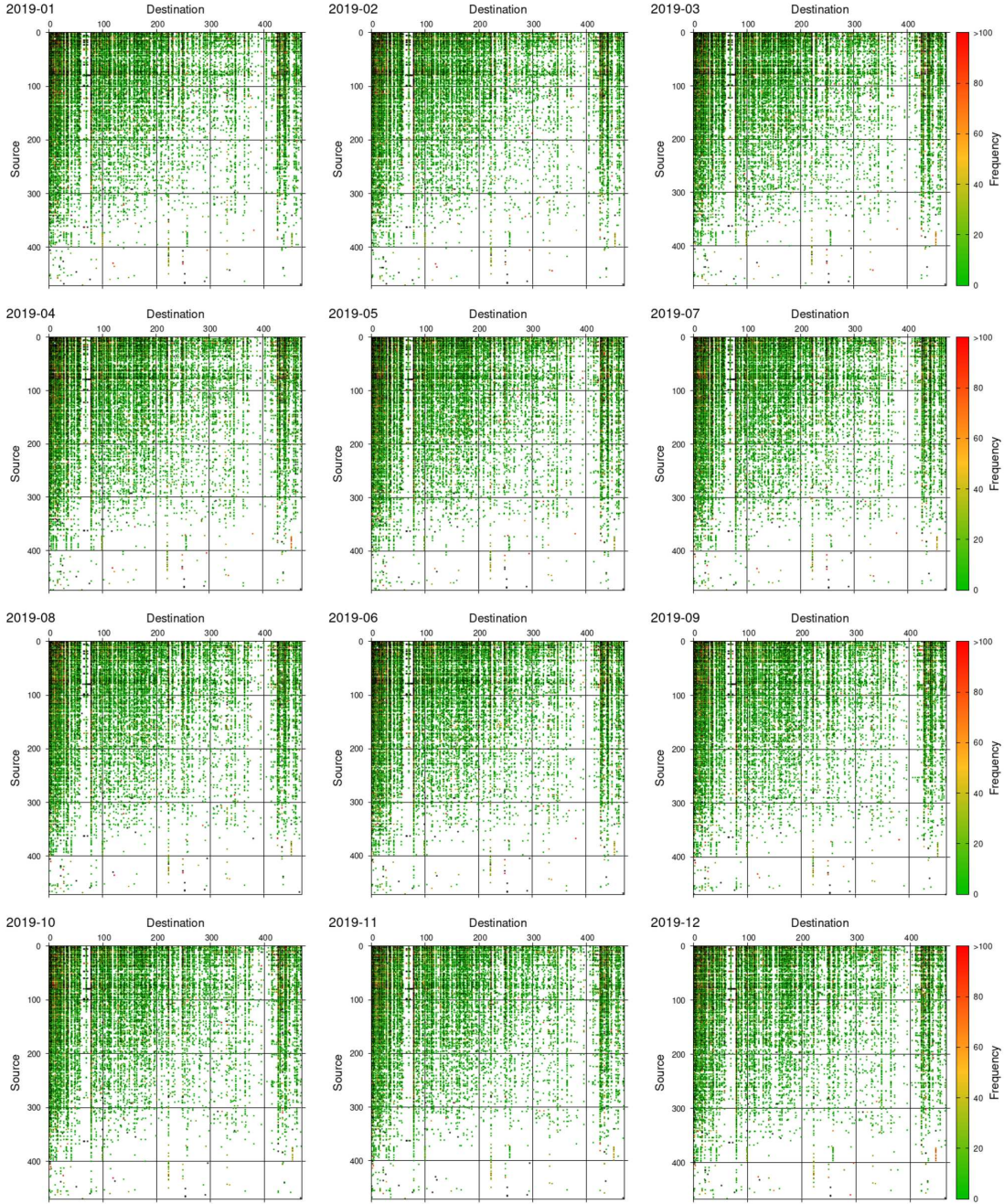}
  \caption{%
    Adjacent matrices for all the networks $G_t$ in
    \tabref{tab:nw_basic}, or equivalently $X_{sd}$ given by
    \eqref{eq:Xandf}. Colors show the strength of links,
    expressed by the frequencies.}
  \label{fig:adj}
\end{figure}

\figref{fig:adj} illustrates the adjacency matrices for all the networks
$G_t$ in \tabref{tab:nw_basic}, or equivalently the non-negative
matrix $X_{sd}$ given by \eqref{eq:Xandf}.

\clearpage
\section{Topic Model of Latent Dirichlet Allocation (LDA)}\label{sec:lda}

Topic models in natural language processing and machine learning are
probablistic models for how terms appear in documents in a
given corpus. Assumed is the term frequency, irrespective of
the sequential order of term occurences; such an assumption is
often called ``bag-of-words''. A topic is a latent variable
in the model for explaining term frequency occurrences. Latent
Dirichelt allocation (LDA) \cite{blei2003tm} is such a model widely
used in a variety of applications. See
\cite{griffiths2004fst,blei2009tm,grun2011tm} and references therein
for introduction and applications. In this appendix, we briefly
summarize the topic model only for the purpose of explicitly
relating it with non-negative matrix factorization (NMF).

$D$ documents in a corpus (a collection of documents) are given
with a vocaburary of $V$ terms (different words). Document $d$
($d=1,\ldots,D$) comprises $N_d$ words, possibly with duplication,
each denoted by $w_{dn}$ ($n=1,\ldots,N_d$). $K$ topics are assumed to
be present for the given corpus.  A topic is a latent or hidden
variable giving a probability distribution for term occurrences. If a
document is an article on sports, terms like ``swimming'' and
``gymnastics'' are likely to occur, while ``market'' and ``employee''
will be unlikely.

Consider the example of an article on ``influence of hosting Olylmics
to economy'' in a corpus of newspaper. It would be natural to consider
that the article belongs to two topics of sports and economy. In topic
models, documents are not assumed to belong to a single topic, but to
simultaneously belong to multiple topics, and the topics vary among
documents. Topic models aim at modeling such mixed membership.

Document $d$ has a \textit{topic distribution}, which is a multinomial
distribution with the parameters:
\begin{equation}
  \label{eq:def_theta}
  \bm{\theta}_d=(\theta_{d1},\ldots,\theta_{dK})\ ,
\end{equation}
where $\theta_{dk}=p(k\,|\,\bm{\theta}_d)$ is the probability that a topic
$k$ is assigned to the document $d$, satisfying
\begin{equation}
  \label{eq:cond_theta}
  \theta_{dk}\geq 0 \quad\text{ and }\quad
  \sum_{k=1}^K\theta_{dk}=1\ .
\end{equation}
Then a topic $z_{dn}\in\{1,\ldots,K\}$ is assigned to each word
$w_{dn}$ for $n=1,\ldots,N_d$. Each topic $k$ is a \textit{term distribution},
which is also a multinomial distribution with the parameters:
\begin{equation}
  \label{eq:def_phi}
  \bm{\phi}_k=(\phi_{k1},\ldots,\phi_{kV})\ ,
\end{equation}
where $\phi_{kv}=p(v\,|\,\bm{\phi}_k)$ is the probability that
a term $v$ occurs, satisfying
\begin{equation}
  \label{eq:cond_phi}
  \phi_{kv}\geq 0 \quad\text{ and }\quad
  \sum_{v=1}^V\phi_{kv}=1\ .
\end{equation}

The topic model of LDA can be most readily understood by looking at
how it generates words in documents as follows.
\begin{enumerate}\renewcommand{\labelenumi}{\arabic{enumi}.}
\item For each topic $k=1,\ldots,K$, generate the parameters
  $\bm{\phi}_k$ for the term distribution by
  \begin{equation}
    \label{eq:gen_phi}
    \bm{\phi}_k\sim\text{Dir}(\beta)\ ,
  \end{equation}
  where $\sim$ reads that a variable is realized as a sample
  from the distribution on the right-hand side, and
  $\text{Dir}(\cdot)$ is the Dirichlet distribution\footnote{%
    Dirichlet distribution is defined by
    \begin{equation}
      \text{Dir}(\bm{x}\,|\,\bm{\xi})=
      \frac{\Gamma\left(\sum_{i'=1}^I\xi_{i'}\right)}{\prod_{i''=1}^I\Gamma(\xi_{i''})}\times
      \prod_{i=1}^I x_i^{\xi_i-1}\ ,\nonumber
    \end{equation}
    where $I$ is the number of variables and it is assumed that
    $x_i\geq 0$ for all $i$ and $\sum_{i=1}^I x_i=1$.
    $\Gamma(\cdot)$ is the gamma function defined by
    $\Gamma(u)=\int_0^\infty t^{u-1}\,e^{-t}dt$ for $u>0$.
    Dirichlet distribution is a generalization of Beta distribution
    (the case $I=2$).}.
  $\bm{\beta}=(\beta_1,\ldots,\beta_V)$ is a set of hyper-parameters with
  $\beta_i>0$. 
\item Now for each document $d=1,\ldots,D$,
  \begin{enumerate}\renewcommand{\labelenumii}{(\alph{enumii})}
  \item generate the parameters $\bm{\theta}_d$ for the topic distribution by
    \begin{equation}
      \label{eq:gen_theta}
      \bm{\theta}_d\sim\text{Dir}(\bm{\alpha})\ ,
    \end{equation}
    where $\bm{\alpha}=(\alpha_1,\ldots,\alpha_K)$ is a set of hyper-parameters with
    $\alpha_k>0$.
  \item For each word $w_{dn}$ ($n=1,\ldots,N_d$)
    \begin{enumerate}\renewcommand{\labelenumiii}{(\roman{enumiii})}
    \item genrate a topic by
      \begin{equation}
        \label{eq:gen_topic}
        z_{dn}\sim\text{Cat}(\bm{\theta}_d)\ ,
      \end{equation}
      where $\text{Cat}(\cdot)$ is the categorical distribution\footnote{%
        Categorical distribution is a multinomial distribution for
        a single trial (a roll of a dice), i.e.
        \begin{equation}
          \label{eq:def_cat}
          \text{Cat}(v\,|\,\bm{\phi})=\phi_v\ ,\nonumber
        \end{equation}
        where $v\in\{1,\ldots,V\}$ and $\bm{\phi}=(\phi_1,\ldots,\phi_V)$
        with $\phi_v\geq 0$ for all $v$ and $\sum_v\phi_v=1$.}, and then
    \item genrate a word from the vocaburary by
      \begin{equation}
        \label{eq:gen_word}
        w_{dn}\sim\text{Cat}(\bm{\phi}_{k=z_{dn}})\ .
      \end{equation}
    \end{enumerate}
  \end{enumerate}
\end{enumerate}
Note that a topic is assigned to each occurence of word $w_{dn}$
at the location $n$ of the document $d$, depending on how the document
$d$ has a mixture of topics, but is not associated with each term $v$
in the vocaburary of $V$ terms. In the previous example, the term
``swimming'' occurs with a relatively high probability if the topic of
sports is assigned at the location of occurence, but the term
will not be likely to occur if the topic of economy is assigned
at the same location.

Given the number of topics, $K$, and the hyper-parameters,
$\bm{\beta}$ and $\bm{\alpha}$, the model $\mathcal{M}$ can be
specified by the parameters $\{\bm{\theta}_d\}_{d=1,\ldots,D}$ and
$\{\bm{\phi}_k\}_{k=1,\ldots,K}$, while the data $\mathcal{D}$
comprises $D$ documents, each of which is a set of words
$\{w_{dn}\}_{n=1,\ldots,N_d}$ for $d=1,\ldots,D$.
In the Bayesian framework of
\begin{equation}
  \label{eq:lda_bayes}
  p(\mathcal{M}\,|\,\mathcal{D})
  =\frac{p(\mathcal{D}\,|\,\mathcal{M})\cdot p(\mathcal{M})}{p(\mathcal{D})}\ .
\end{equation}
the prior distribution $p(\mathcal{M})$ is given by the Dirichlet
distributions of \eqref{eq:gen_phi} and \eqref{eq:gen_theta}. It is
easy to show that the likelihood $p(\mathcal{D}\,|\,\mathcal{M})$
can be written by
\begin{equation}
  \label{eq:lda_ll}
  \log p(\mathcal{D}\,|\,\mathcal{M})=
  \sum_{d=1}^D\sum_{n=1}^{N_d}\log\sum_{k=1}^K \theta_{dk}\cdot\phi_{kw_{dn}}\ ,
\end{equation}
which follows from the assumption of bag-of-words and the independence
of documents. Estimation of parameters by maximum likelihood or
Bayesian frameworks has a difficulty due to the log of sum in
\eqref{eq:lda_ll}, so there are a variety of technical methods such as
EM (expectation and maximization), variational Bayes, and MCMC (Markov
chain Monte Carlo) algorithms (see
\cite{griffiths2004fst,blei2009tm,grun2011tm} and references therein).

For our purpose, let us turn out attention to how a word is chosen
from the $V$ terms in the vocaburary in the $D$ documents, given
the model parameters. Looking at \eqref{eq:gen_topic} and
\eqref{eq:gen_word}, one can see that the probability that a term
$v$ is chosen as a word in the document $d$, denoted by
$\lambda_{dv}$, is given by
\begin{equation}
  \label{eq:lda_nmf}
  \lambda_{dv}=\sum_{k=1}^K\theta_{dk}\cdot\phi_{kv}\ ,
\end{equation}
where it is assumed that
\begin{equation}
  \label{eq:cond_lamb}
  \lambda_{dv}\geq 0 \quad\text{ and }\quad
  \sum_{v=1}^V\lambda_{dv}=1\ .
\end{equation}
(Note that \eqref{eq:lda_nmf} is the factor in the likelihood
\eqref{eq:lda_ll}). In terms of matrices, the $D\times V$ matrix,
denoted by $\bm{\Lambda}$, with elements $\lambda_{dv}$ is represented
by a product of two matrices; the $D\times K$ matrix $\bm{\Theta}$
with elements $\theta_{dk}$ times the $K\times V$ matrix $\bm{\Phi}$
with elements $\phi_{kv}$. Note that it is usually the case that
$K\ll V$ and $K\ll D$. One can immediately see that \eqref{eq:lda_nmf} is
actually a non-negative matrix factorization (NMF) of $\bm{\Lambda}$
by two generally much smaller matrices:
\begin{equation}
  \label{eq:lda_nmf_mat}
  \bm{\Lambda}=\bm{\Theta}\cdot\bm{\Phi}
\end{equation}
under the conditions of \eqref{eq:cond_theta} and \eqref{eq:cond_phi},
from which \eqref{eq:cond_lamb} follows.

A document-term matrix is the $D\times V$ matrix, the rows and columns
of which correspond to the documents and terms respectively, and each
element is the frequency (or any other measure that represents some
intensity) for the term occurences. One can regard the document-term
matrix as a realization in the sense that each element at column $v$
and row $d$ represents how the term $v$ was realized in the document
$d$ according to the categorical (multinomial) distribution with the
parameter $\lambda_{dv}$. See the main text in \secref{sec:nmf_prob}.

How to determine the number of topics, $K$, is important.
There have been a number of works on this issue.
A class of studies \cite{cao2009dbm,arun2010fnn,deveaud2014ael} is
based on some measure to compute similarities of extracted topics,
called coherence. The idea is that an optimal value of $K$ is the
point where the overall dissimilarity among topics achieves a maximum,
because extracted topics should be different to a certain degree.
Other approaches include the idea of perplexity
\cite{griffiths2004fst}, hierarchical Dirichlet process
\cite{teh2006hdp}, and so forth.  It is beyond the present paper to
review these approches. Interested readers are guided to look at them
and reviews \cite{blei2009tm,grun2011tm} with references therein.  In
the main text, we use the idea of coherence
\cite{cao2009dbm,arun2010fnn,deveaud2014ael}, which we find to work
well for our purpose.

\clearpage
\section{Optimal Number of NMF Components}\label{sec:optnum}

In \secref{sec:nmf_prob}, we showed that the matrix decomposition of
NMF can be regarded as a probablistic model, which is equivalent to a
model of probabilistic latent semantic analysis (PLSA) in natural
language processing, in particular, latent Dirichlet allocation
(LDA). The number of NMF components is such an analogous parameter as
the number of topics in the latter. Then, as a bonus, one can employ
the method of estimating the number of topics to our problem of
detemining the number of NMF components, $K$. We found that three
different methods in the literature
\cite{cao2009dbm,arun2010fnn,deveaud2014ael} give a consistent result
for the optimal value of $K$, and also that one of them
\cite{arun2010fnn} is relatively stable in giving the optimal (see
\figref{fig:detK1909}).

\begin{figure}[tb]
  \centering
  \includegraphics[width=0.98\linewidth]{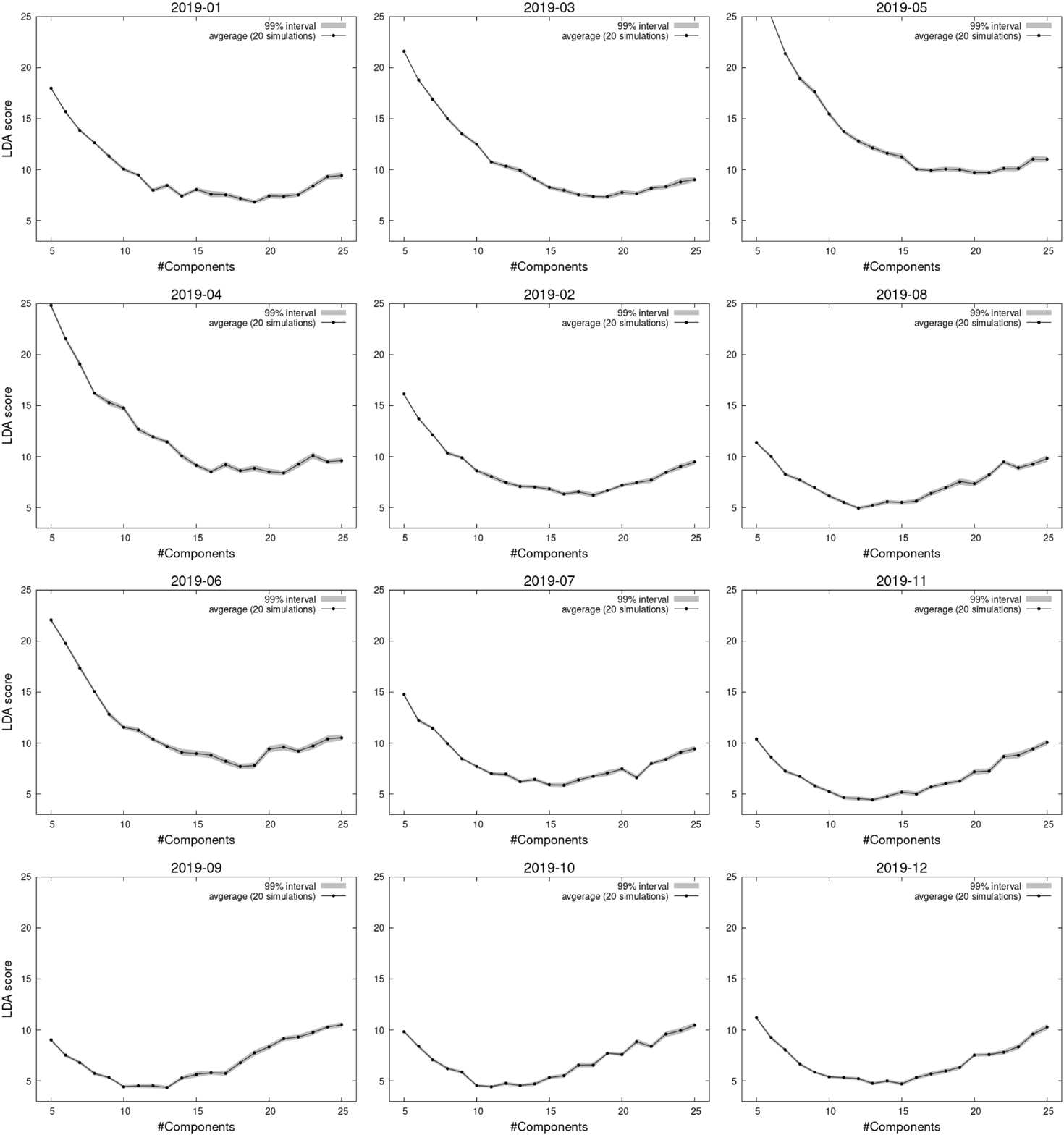}
  \caption{%
    Determining $K$, the number of NMF components, by minimizing the
    measure of coherence \cite{arun2010fnn}. We performed Monte Carlo
    simulations with 20 runs for each number of components. Averages
    (points) and 99\% level (gray band, narrow) calculated from
    standard errors are drawn. Data: all months in the year 2019.}
  \label{fig:fig_ldat_all}
\end{figure}

In \figref{fig:fig_ldat_all}, we summarize the same result for the
data in all the other months in the same year. We found that the
optimal number $K$ is not volatile in its variation among different
months, and also small in its value ranging around 11--18.
\tabref{tab:optnum} shows the optimal values for all the months
of the year 2019.

\begin{table}[tbh]
  \centering
  \caption{%
    Optimal values for the number of NMF components, $K$,
    estimated from \figref{fig:fig_ldat_all}}
  \begin{tabular}{cc}
    \toprule
    Month & Optimal $K$ \\
    \midrule
    01 & 14 \\
    02 & 16 \\
    03 & 18 \\
    04 & 16 \\
    05 & 16 \\
    06 & 18 \\
    07 & 15 \\
    08 & 12 \\
    09 & 13 \\
    10 & 11 \\
    11 & 13 \\
    12 & 13 \\
    \bottomrule
  \end{tabular}
  \label{tab:optnum}
\end{table}
  
\end{document}